\documentclass[useAMS,fleqn,usenatbib]{mnras}

\usepackage{newtxtext,newtxmath}
\usepackage{amsmath}
\usepackage[utf8]{inputenc}
\usepackage{physics}
\usepackage[utf8]{inputenc}
\usepackage{changepage}
\usepackage{afterpage}
\usepackage{floatpag}
\usepackage{transparent}
\usepackage{placeins}
\usepackage{physics}
\usepackage{booktabs}

\usepackage[T1]{fontenc}

\DeclareRobustCommand{\VAN}[3]{#2}
\let\VANthebibliography\thebibliography
\def\thebibliography{\DeclareRobustCommand{\VAN}[3]{##3}\VANthebibliography}

\usepackage{graphicx}	
\usepackage{amsmath}	
\usepackage{ulem}
\usepackage{cancel}
\usepackage{xcolor}
\usepackage{grffile}
\usepackage{url}
\usepackage{comment}
\usepackage{natbib}
\usepackage{enumerate}
\usepackage{verbatim}
\usepackage{float}
\usepackage{placeins}
\usepackage{soul}

\newcommand\sect{Section}

\newcommand\tabl{Table}
\newcommand\eqn{Eq.~}
\newcommand\eqns{Eqs.~}
\newcommand\fig{Fig.~}

\newcommand{\msun}{\mathrm{M}_{\odot}}
\newcommand\wlmass{M_{\rm{WL}}}
\newcommand\truemass{M_{\rm{true}}}

\newcommand\mfh{M_{500}}

\newcommand\rfh{r_{500}}
\newcommand\rmin{r_{\rm{min}}}
\newcommand\rmax{r_{\rm{max}}}
\newcommand\rmis{R_{\rm{mis}}}
\newcommand\meanrmis{\overline{R}_{\rm{mis}}}

\newcommand\bias{b}
\newcommand\logbias{\log{\bias}}

\newcommand\munormal{\mu_{\bias}}
\newcommand\sigmanormal{\sigma_{\bias}}
\newcommand\mulognormal{\mu_{\logbias}}
\newcommand\sigmalognormal{\sigma_{\logbias}}

\newcommand\overcorrection{\tau}
\newcommand\overcorrectionnormal{\tau_{\bias}}
\newcommand\overcorrectionlognormal{\tau_{\logbias}}

\newcommand\wnormal{w_{\bias}}
\newcommand\wlognormal{w_{\logbias}}

\newcommand\nonrandommiscentring{\rm{(s)}}
\newcommand\randommiscentring{\rm{(r)}}

\newcommand\cfit{\rm{(c)}}
\newcommand\mfit{\rm{(m)}}

\newcommand\zlens{z_{\mathrm{l}}}
\newcommand\zsource{z_{\mathrm{s}}}
\newcommand\esource{\epsilon_{\mathrm{s}}}

\newcommand{\appropto}{\mathrel{\vcenter{
  \offinterlineskip\halign{\hfil$##$\cr
    \propto\cr\noalign{\kern2pt}\sim\cr\noalign{\kern-2pt}}}}}

\title[Weak lensing mass bias and miscentring]{Weak lensing mass bias and the alignment of center proxies}

\author[Sommer et al.]{Martin W. Sommer$^{1}$\thanks{E-mail: mnord@astro.uni-bonn.de (MWS)
},
Tim Schrabback$^{1,2}$,
Antonio Ragagnin$^{3,4,5}$
and Robert Rockenfeller$^{6}$
\\
$^{1}$Argelander-Institut f\"{u}r Astronomie, Auf dem H\"ugel 71, D-53121 Bonn, Germany \\
$^{2}$Institut für Astro- und Teilchenphysik, Universität Innsbruck, Technikerstr. 25/8, 6020 Innsbruck, Austria \\ 
$^{3}$Dipartimento di Fisica e Astronomia "Augusto Righi", Alma Mater Studiorum Università di Bologna, via Gobetti 93/2, I-40129 Bologna, Italy \\ 
$^{4}$INAF - Osservatorio Astronomico di Trieste, via G.B. Tiepolo 11, I-34143 Trieste, Italy \\
$^{5}$IFPU - Institute for Fundamental Physics of the Universe, Via Beirut 2, I-34014 Trieste, Italy \\
$^{6}$Mathematisches Institut, Universität Koblenz, Universtitätsstr. 1, D-56070 Koblenz, Germany
}

\date{Accepted XXX. Received YYY; in original form ZZZ}

\pubyear{2021}

\begin{document}
\label{firstpage}
\pagerange{\pageref{firstpage}--\pageref{lastpage}}
\maketitle

% Abstract of the paper
\begin{abstract}
Galaxy cluster masses derived from observations of weak lensing suffer from a number of biases affecting the accuracy of mass-observable relations calibrated from such observations. In particular, the choice of the cluster center plays a prominent role in biasing inferred masses. In the past, empirical miscentring distributions have been used to address this issue. Using hydro-dynamical simulations, we aim to test the accuracy of weak lensing mass bias predictions based on such miscentring distributions by comparing the results to mass biases computed directly using intra-cluster medium (ICM)-based centers from the same simulation. We construct models for fitting masses to both centered and miscentered Navarro-Frenk-White profiles of reduced shear, and model the resulting distributions of mass bias with normal and log-normal distributions. 
We find that the standard approach of using miscentring distributions leads to an over-estimation of cluster masses at levels of between 2\% and 6\% when compared to the analysis in which actual simulated ICM centers are used, even when the underlying miscentring distributions match in terms of the miscentring amplitude. We find that neither log-normal nor normal distributions are generally reliable for approximating the shapes of the mass bias distributions, regardless of whether a centered or miscentered radial model is used. 
\end{abstract}

\begin{keywords}
gravitational lensing: weak -- galaxies: clusters: general
\end{keywords}

\section{Introduction}
\label{sec:intro}

The distribution and abundance of clusters of galaxies throughout the Universe are sensitive to both the 
geometry of the Universe and the growth rate of primordial density fluctuations (e.g.~\citealt{2001ApJ...553..545H}), and thus constitute a powerful tool for investigating the dark energy equation of state and other aspects of the so-called standard cosmological model \citep[for a review see, e.g.,][]{2011ARA&A..49..409A}. The accurate calibration of mass-observable relations is in turn a critical factor in constraining cosmological models from the mass-redshift distribution of clusters of galaxies in large-scale surveys. Weak lensing (henceforth also WL) observations provide a way of measuring masses of galaxy clusters in a direct way through distortions in the images of background galaxies.

Such masses, however, suffer from a number of systematic biases and uncertainties, which at present tend to 
dominate the systematic error budget in mass-observable relations calibrated on weak lensing measurements (e.g. \citealt{2014MNRAS.439...48A, 2014MNRAS.440.2077M,2016A&A...594A..24P,2018MNRAS.474.2635S, 2019MNRAS.483.2871D,2019ApJ...878...55B,2019MNRAS.482.1352M,2021MNRAS.505.3923S,2022A&A...668A..18Z}). We divide these systematics into two categories which we label (i) \textit {observational} and (ii) \textit{modeling related}.  
In the first category, we include systematics directly related to observational data. These include the selection of background galaxies (contamination from the inadvertent inclusion of cluster members), systematics in the shape measurements and lensing efficiency estimates of said galaxies, and systematics pertaining to various observational correction terms (boost and magnification corrections). The second category includes systematics related to the accuracy of WL mass modeling. In a sense, these are the systematics that remain under the assumption that the WL data have been perfectly calibrated. We subdivide this group into four parts. First, for any galaxy cluster, there will be projection of mass along the line of sight (correlated and uncorrelated large-scale structure, or LSS). Second, in part due to limited signal-to-noise ratios of observations, masses are typically modeled using spherical (or ellipsoidal) models (e.g. \citealt{2016MNRAS.457.1522A,2019MNRAS.483.2871D}), whereby the effects of triaxiality cannot be fully captured. Third, the use of parametric models of radial mass density introduce a further model-inherent bias, and fourth, the choice of the coordinate at which the radial model is centered can over- or underestimate the mass of a cluster. 

The understanding of systematic effects directly related to the accuracy in WL mass modeling (category (ii)) ideally need to match not only the statistical uncertainties of masses in upcoming galaxy clusters surveys, but also the level of systematic uncertainty from observational effects (category (i)). 
\cite{2019MNRAS.488.2041G} estimated a combined contributions to the systematic error budget of the absolute mass calibration from the latter at around one per cent for the \textit{Euclid}\footnote{http://sci.esa.int/euclid/} \citep{2011arXiv1110.3193L} and Rubin Observatory Legacy Survey of Space and Time\footnote{https://www.lsst.org/} \citep[LSST,~][]{2019ApJ...873..111I} surveys. Currently, quoted systematic uncertainties on mass modeling bias range from a few to several per cent (e.g. \citealt{2019MNRAS.483.2871D,2021MNRAS.505.3923S,2021MNRAS.507.5671G,2022A&A...668A..18Z,2023MNRAS.tmp..949C}).
Significant efforts have been made towards quantifying the weak lensing mass bias distribution, both from N-body dark matter-only (DMO) simulations  \citep{2011ApJ...740...25B,2011MNRAS.414.1851O,2012MNRAS.421.1073B} as well as from hydrodynamic simulations \citep{2017MNRAS.465.3361H,2018MNRAS.479..890L}, mostly considering cases in which the gravitational center of a halo is known to arbitrary accuracy (for a summary, see the introduction of \citealt{2022MNRAS.509.1127S}). 

The abundance of galaxy clusters as a function of redshift is dominated by gravitational effects, and observed masses are often poorly matched to theoretical masses due to uncertainties in the gravitational center of galaxy clusters in the sense of the bottom of the gravitational potential well. In simulations, this center is closely approximated by the position of the most bound particle, while in observations, it often coincides well with the brightest cluster galaxy (BCG). However, the latter is not necessarily a suitable center proxy in general, as in many cases the BCG is far from the gravitational center (e.g. \citealt{2012ApJ...757....2G}). In particular, merging clusters may have more than one BCG candidate leading to a bi-modal miscentring distribution.

Apart from using the BCG, the center coordinate can be chosen in different ways: it may be modeled directly in conjunction with the shear profile, or one may use the peak in the weak lensing convergence (derived from the reduced shear), as investigated in some detail by \citet{2022MNRAS.509.1127S}. It may also be derived from the peak or centroid of an observable of the intra-cluster medium (ICM), such as the Compton-Y distortion (Sunyaev-Zeldovich effect, also SZE, \citealt{1970CoASP...2...66S,1980ARA&A..18..537S}) or the X-ray brightness.

We denote the offset between the gravitational center and the chosen center of a cluster with the term \textit{miscentring}. For most center proxies, miscentring effects result in underestimated masses on average. This problem has been investigated using distributions of miscentring with random orientations (e.g., \citealt{2011ApJ...740...25B, 2019MNRAS.483.2871D,2021MNRAS.507.5671G,2021MNRAS.505.3923S, 2022MNRAS.509.1127S}). While such an approach has some merit in the sense that it accounts for random deviations from the gravitational center, it does not account for inherent (directional) correlations between the center proxy and weak-lensing shear measurements. This is the central issue under investigation in this work, in particular regarding proxies derived from the ICM.

We make use of the hydrodynamic simulations to generate lensing and SZE images of hundreds of galaxy clusters at redshift $z=0.67$. In a first step, we compare the weak-lensing mass bias arising when center proxies taken from the ICM (specifically, center proxies from the Sunyaev-Zeldovich effect and X-ray bremsstrahlung) are used, to the corresponding bias arising when random positions from the same radial miscentring distribution are used. In a second step, we add another miscentring distribution, approximately mimicking the effects of SZE measurement noise and primary CMB anisotropies. In addition to comparing the resulting mass bias distributions (not to be confused with the miscentring distributions) in terms of their averages and dispersions, we also investigate these bias distributions in terms of how well they can be described by normal and log-normal probability distributions. We make use of both centered and miscentered azimuthally symmetric models of mass density.

This paper is structured as follows. In \sect~\ref{sec:meth} we briefly summarize the underlying theory of weak lensing, introduce our sample of simulated clusters and describe how we generate miscentring distributions. We describe an often used radial model, and discuss how the latter can be translated into a miscentered model, and finally outline a simple framework for judging how well a mass bias distribution can be said to be consistent with a normal or a log-normal distribution. In \sect~\ref{sec:results}, we test for a mass dependence in the weak lensing mass bias, and investigate how the mass bias distribution changes when using a random miscentring distribution rather than one inferred directly from the simulation (with a specific miscentring for each target, corresponding to the ICM in the simulation). We also test how this shift in the mass bias distribution depends upon the radial range of weak lensing data. We discuss the implications of our results in \sect~\ref{sect:discussion} and offer our conclusions in \sect~\ref{sec:conclusion}. 

Throughout the paper, we use the flat $\Lambda$CDM cosmological model associated with the Magenticum simulations (Hubble parameter $h = 0.703$, total mass density $\Omega_{\rm{m}} = 0.272$, dark energy density $\Omega_\Lambda=0.728$). We quantify masses of galaxy clusters in terms of the over-density parameter $\Delta$. With $r_{\Delta}$ the radius within which the mean density is equal to $ \rho_{\rm{crit}} \Delta$, where $\rho_{\rm{crit}}$ is the critical density of the Universe at the redshift of the cluster, $M_{\Delta}$ is then the mass inside $r_{\Delta}$. In particular, halo masses considered in this work are $\mfh$
unless otherwise noted. In general, we drop the index "500" when discussing true vs. measured masses. Units of distance are in physical, not comoving, Megaparsec (Mpc). We take $\log(x)$ to mean the natural logarithm of $x$. As we introduce a large number of specific quantities, a summary of non-standard notation is given in   Appendix~\ref{sec:appendix:notation}. 

\section{Methods}
\label{sec:meth}

\subsection{Weak lensing formalism}
\label{sec:meth:wl}

Gravitational lensing by a foreground lens (such as a cluster of galaxies) at redshift $\zlens$ introduces a distortion in the images of a background (``source'') galaxy at redshift $\zsource$. Formally, the shape distortion is described by the transformation matrix
\begin{equation}
\label{eq:atransf}
  \mathbf{A(\btheta)} =
  \left[ {\begin{array}{cc}
    1 - \kappa(\btheta) - \gamma_1(\btheta) & - \gamma_2(\btheta) \\
    - \gamma_2(\btheta) & 1 - \kappa(\btheta) + \gamma_1(\btheta) \\
  \end{array} } \right],
\end{equation}
where in the following we shall drop the (two-dimensional) positional argument $\btheta$ and consider it implicit. The (unobservable) shear $\gamma = \gamma_1 + \rm{i}\gamma_2$ is an anisotropic distortion, while the isotropic distortion is described by the convergence $\kappa(\btheta)=\Sigma(\btheta) / \Sigma_{\mathrm{crit}}$, which
is the surface mass density $\Sigma(\btheta)$ expressed in units of the critical surface density $\Sigma_{\text{crit}}$. The latter is defined through
\begin{equation}
\frac{1}{\Sigma_{\text{crit}}} = \frac{4 \pi G}{c^2} D_{\mathrm{l}}\beta
\end{equation}
where $c$ is the speed of light, $G$ is the gravitational constant and the lensing efficiency $\beta$ is defined by
\begin{equation}
\beta = \frac{D_{\text{ls}}}{D_{\text{s}}} H(\zsource-\zlens),
\end{equation}
where $D_{\mathrm{s}}$, $D_{\mathrm{l}}$ and $D_{\mathrm{ls}}$ are the angular diameter distances between the observer and the source, the observer and the lens, and the lens and the source, respectively. The Heaviside step function, $H(x)$, is equal to one for positive values of $x$, and zero otherwise (the latter corresponding to the fact that lensing does not occur unless the source is behind the lens).

By the tranformation matrix of \eqn(\ref{eq:atransf}), a circle is translated into an ellipse.
In the limit of weak lensing ($\kappa \ll 1$), the anisotropic component of the shape distortion is characterized by the reduced shear $g=g_1+\text{i}g_2$, given by
\begin{equation}
    g = \frac{\gamma}{1-\kappa}
\end{equation}
(see, e.g. \citealt{2015RPPh...78h6901K}, for a more detailed account). For $|g| \leq 1$, the reduced shear can be estimated from the ensemble-averaged observed ellipticities\footnote{We define ellipticity as $\epsilon = (a-b)/(a+b) \cdot \mathrm{e}^{\mathrm{2i\phi}}$ for elliptical isophotes with minor-to-major axis ratio $b/a$ and
position angle $\phi$.} $ \epsilon = \epsilon_1 + \mathrm{i} \epsilon_2$, as \citep{1997A&A...318..687S}
\begin{equation}
    \epsilon = \frac{\esource + g}{1 + g^*\esource},
\end{equation}
where $g^*$ denotes the complex conjugate of the reduced shear, and $\esource$ is the intrinsic complex ellipticity of a source galaxy. Because of the intrinsic ellipticities $\esource$, $g$ is not identical to $\epsilon$. However, assuming that the source galaxies have no preferred orientation, the expectation value of $\esource$ vanishes ($ \langle \esource \rangle = 0$), and it holds that $\langle \epsilon \rangle = g$, that is, the ellipticity is an unbiased estimator of the reduced shear.

While the dispersion of intrinsic ellipticities (shape noise) is distinct from uncertainties in measured ellipticities (shear noise), both components are typically combined for the purpose of creating mock weak lensing data from simulations. \citet{2022MNRAS.509.1127S} derived the weak lensing mass bias distribution in the presence of noise, requiring a Bayesian framework and an \textit{a priori} model of the distribution. In that work, it was found that log-normal models of the bias are independent of the noise level, given that the underlying distribution is in fact log-normal. In this work, we assume this to be true also for any other underlying model, as there is no particular reason to suspect that there should be a noise dependence in the bias that is specific to a certain distribution. This, in turn, allows us to inspect bias distributions directly, and compare them to theoretical distributions.

The shear, reduced shear and ellipticity can be decomposed into tangential (subscript $t$) and cross (subscript $\mathsf{x}$) components through 
\begin{equation}
\label{eq:shearxt}
\begin{array}{lcl}
      {(\cdot)}_{\mathrm{t}} & = & - {(\cdot)}_1 \cos{(2 \alpha)}  - {(\cdot)}_2 \sin{(2 \alpha)} \\
      {(\cdot)}_{\times} & = & + {(\cdot)}_1 \sin{(2 \alpha)} - {(\cdot)}_2 \cos{(2 \alpha)},
\end{array}
\end{equation}
where ${(\cdot)}$ denotes any of $g$, $\gamma$ and $\epsilon$, and $\alpha$ is the azimuthal angle with respect to a chosen center. The inverse of these relations is given by
\begin{equation}
\label{eq:shear12}
\begin{array}{lcl}
      {(\cdot)}_1 & = & - {(\cdot)}_{\mathrm{t}} \cos{(2 \alpha)} + {(\cdot)}_{\times} \sin{(2 \alpha)} \\
      {(\cdot)}_2 & = & - {(\cdot)}_{\mathrm{t}} \sin{(2 \alpha)} - {(\cdot)}_{\times} \cos{(2 \alpha)}.
\end{array}
\end{equation}

For an azimuthally symmetric or azimuthally averaged projected mass distribution, the cross shear term vanishes, while the tangential shear as a function of projected radius $r$ can be written as
\citep[e.g.][]{1995ApJ...449..460K,2000ApJ...534...34W}
\begin{equation}
    \gamma_{\mathrm{t}}(r) =  {\bar{\kappa}}(<r) - {\bar{\kappa}}(r),
\end{equation}
where ${\bar{\kappa}}(<r)$ is the mean convergence inside radius $r$, and ${\bar{\kappa}}(r)$ is the azimuthally averaged convergence at radius $r$.
Equivalently, in terms of the surface mass density,
\begin{equation}
\label{eq:gammafrommass}
    \gamma_{\mathrm{t}}(r) = \frac{\overline{\Sigma}(<r)-\overline{\Sigma}(r)}{\Sigma_{\rm{crit}}}.
\end{equation}

\subsection{Sample and images}
\label{sec:meth:sample}

From the Magneticum Pathfinder Simulation \citep{2016MNRAS.463.1797D}, we select the 275 most massive galaxy clusters from snapshot 22, at redshift $z=0.67$, of the box2b-hr simulation box. This simulation has a volume of $640^3 h^{-3}$ Mpc and $2 \times 2880^3$ particles. The mean and median mass of the sample are $M_{500} = 1.69 \times 10^{14} h^{-1} \msun$ and $M_{500} = 1.45 \times 10^{14} h^{-1} \msun$, respectively. 

We project each cluster along three mutually orthogonal axes to yield a final sample of 825 quasi-independent targets, for which we extract X-ray bolometric luminosity, images of thermal and kinematic Compton-Y (Sunyaev-Zeldovich effect) and projected mass density in a $512 \times 512$ pixel grid around each cluster center, with an image size corresponding to 8 times $r_{500}$ and using a depth of 30 Mpc along the line of sight. 

From the projected mass images, we derive the lensing quantities $g_1$, $g_2$ and $\kappa$, described in the previous subsection, using a publicly available software\footnote{github.com/aragagnin/g3read} developed by one of the co-authors. We set a constant lensing efficiency $\beta = 0.3$. This has no direct bearing upon our results, as we work with noiseless shear images. Because shear is a non-local measure, we make the shear images half the size of the original images. 

The X-ray and SZE images are used for deriving quasi-realistic measures of centers, as they would be assigned to galaxy clusters in actual multi-wavelength analyses, as well as for constructing miscentring distributions, as described in the following subsection. While the images of kinematic and thermal SZE are added to yield a total image for each target, the kinematic SZ effect does not impact our results significantly.

The projection of LSS and the effects of triaxiality are not independent. Neighboring halos are generally connected by filaments, and the direction of the major axis of a halo is correlated with the directions to massive neighbors (see, e.g., \citealt{2009ApJ...706..747Z}, and references therein). Such alignments persist out to radii of approximately 100 $h^{-1}$ Mpc from the cluster center \citep{2002A&A...395....1F,2005ApJ...618....1H}, suggesting an optimal integration length of $\sim$200 $h^{-1}$ Mpc ($\pm 100$ $h^{-1}$ Mpc with the halo at zero) for separating correlated and uncorrelated LSS in cosmological simulations. However, \citet{2011ApJ...740...25B} found the WL mass bias distribution to be stable for integration lengths in a range of approximately 30-200 $h^{-1}$ Mpc comoving. Here, we account only for correlated LSS, deriving mock WL data from simulations using similar integration lengths.  

At large radii, correlated matter around the cluster (correlated halos) contributes to the lensing profile (the so-called two-halo term, e.g. \citealt{2000MNRAS.318..203S}, \citealt{2005MNRAS.362.1451M}). Here, we limit the outer radius to 2.2 Mpc (physical) so as to make this term negligible. 

Using ray-tracing through N-body simulations, \cite{2012MNRAS.419.3547D} found that projected large-scale structure, while constituting a major source of systematic uncertainty in weak lensing masses, is not a source of confusion in identifying halo centers. For the purpose of the present work, focusing specifically on the impact of center proxies, it is thus justified to use flat projections from simulations rather than ray tracing.

\subsection{Miscentring distributions}
\label{sec:meth:miscdistr}

We describe here how the SZE and X-ray images, described in the previous subsection, are analyzed in order to derive center measurements around which profiles of tangential reduced shear are constructed. For both SZE and X-ray, we compare our results to reference distributions, which are also based on the Magneticum simulations. The latter, which we henceforth refer to as X-ref and SZ-ref, were described in detail by \cite{2021MNRAS.505.3923S}. We summarize the most important points here. Mock SZE observations were extracted from light cones of the Magneticum simulation, accounting for contributions from CMB anisotropies, a one arcminute resolution and noise filtering. Cluster candidates were identified using an approach adopted for SPT clusters, described in detail by \citet{2009ApJ...701...32S}. The resulting sample of selected clusters was used to characterize both the SZE and X-ray miscentring distributions. Cut-outs of X-ray surface brightness maps were produced at the point of the deepest potential of each halo. The X-ray miscentring distribution was then derived as the distribution of the projected offsets between the peak of the X-ray surface brightness maps and the position of the deepest potential in the halo.

Our measure of miscentring is $\rmis$, defined as the distance in the sky plane between the measured center coordinate and the bottom of the gravitational potential. 

We convolve the SZE images with a one arcminute Gaussian, approximating the resolution of typical ground-based observations. The peaks in the resulting images are taken as the SZE centers, resulting in a miscentring distribution, henceforth SZ-0, that is artificially narrow because primary anisotropies of the cosmic microwave background (CMB), millimeter emission due to dusty galaxies, and instrumental and atmospheric noise components have not been accounted for.

\begin{figure}
\centering
\includegraphics[width=0.97\columnwidth,clip=True,trim={15 7 35 25}]{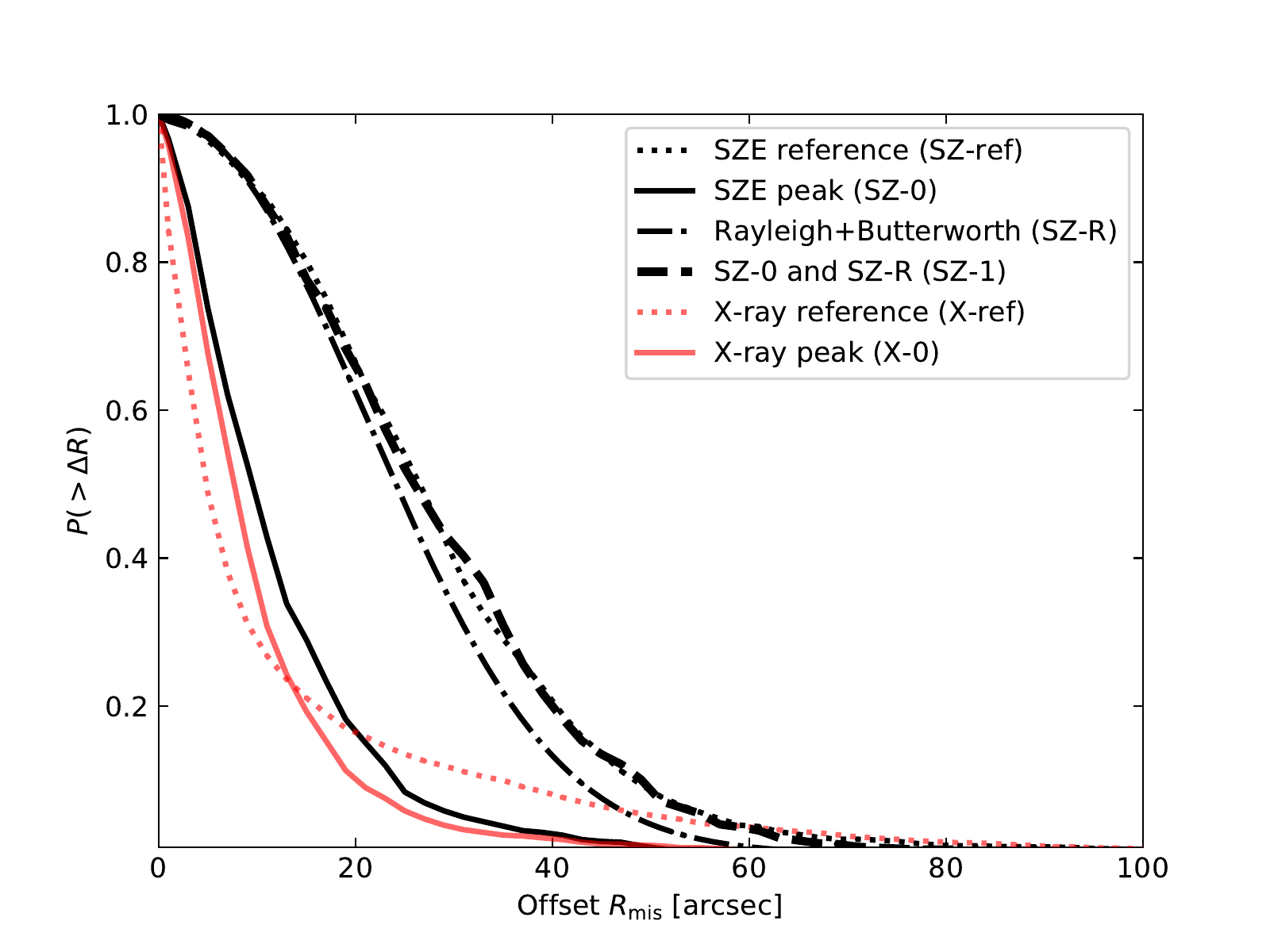} \\
\includegraphics[width=0.97\columnwidth,clip=True,trim={15 7 35 25}]{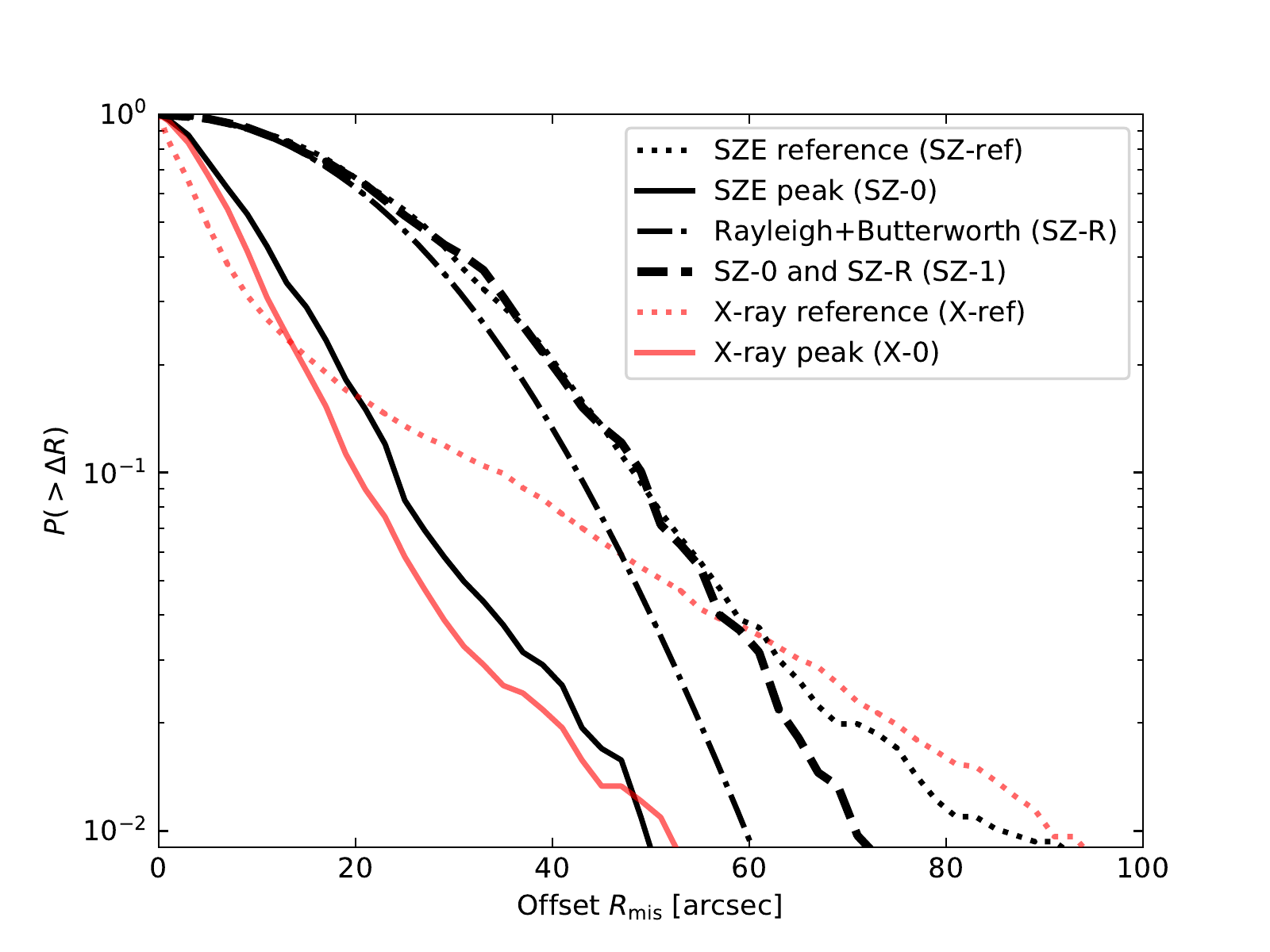}
\caption{\label{fig:miscdistros} Miscentring distributions relevant to this work, in linear scale (top) and log-scale (bottom). The black solid line indicates the miscentring distribution for SZE observations, derived from the Magneticum images with no noise added. To approximately match the reference distribution SZ-ref (black dotted line; see text), we add an extra miscentring, represented by the dash-dotted line, resulting in a net distribution (dashed line) approximating the reference distribution. Also shown are the measured distribution of offsets from X-ray peaks (solid red line) and a corresponding reference distribution (dotted  red line).} 
\end{figure}

In a subsequent step, we attempt to approximately reproduce the SZE reference miscentring distribution, SZ-ref, from SZ-0 by adding miscentrings with random orientations and a miscentring distribution SZ-R with probability density $p(\rmis)$. Specifically, we use a Rayleigh distribution coupled with a first order Butterworth low-pass filter, described by
\begin{equation}
\label{eq:rayleigh}
    p(\rmis) = \frac{\rmis}{\sigma_R^2} \exp\left({\frac{-\rmis^2}{2 \sigma_R^2}}\right) \frac{G_0}{1+\left(\frac{\rmis}{t}\right)^{2}},
\end{equation}
where $\sigma_R = 23 ^{ \prime \prime}$, $t = 46 ^{\prime \prime}$ and $G_0$ is a constant of normalization. We find that SZ-0 coupled with SZ-R closely matches the reference distribution SZ-ref. We see some deviation in the tail of the distribution, affecting a small percentage of the targets, as can be seen in \fig\thinspace\ref{fig:miscdistros}. 

From the X-ray images, we mask point sources (AGN), and iterate on $\rfh$, weighting by the bolometric luminosity, to find the centroids. We also find the peaks after filtering out the AGN with a low-pass filter at a scale of 20 arcseconds, arriving at a consistent miscentring distribution. Because these distributions are both wider than the reference distribution X-ref in the central parts, we do not attempt to reproduce the latter by adding noise to the former.  

\begin{table*}
\centering
\begin{tabular}{r c l l l} 
 \hline
 Label & $\meanrmis$ & miscentring direction & type & Description \\
 & (Mpc) & & & \\
 \hline
 SZ-ref & 0.201 & random & broad  & SZE reference distribution \citep{2021MNRAS.505.3923S}\\
 SZ-0   & 0.092 & actual & narrow & SZE centers from simulation (no added noise) \\
 SZ-0r  & 0.092 & random & narrow & Randomized form of SZ-0 \\
 SZ-R   & 0.179 & & & Tapered Rayleigh distribution (\eqn~\ref{eq:rayleigh}) \\
 SZ-1   & 0.201 & actual & broad & SZ-0 combined with SZ-R \\
 SZ-1r  & 0.201 & random & broad & SZ-0r combined with SZ-R \\
 \hline
 X-ref & 0.090 & random & broad & X-ray reference distribution \citep{2021MNRAS.505.3923S}\\
 X-0   & 0.078 & actual & narrow & X-ray centroids from simulation (no added noise) \\
 X-0r  & 0.078 & random & narrow & Randomized form of X-0 \\
 \hline
\end{tabular}
\caption{Miscentring distributions used in this work. $\meanrmis$ gives the average miscentring radius for each distribution. The miscentring direction 'actual' means taken directly from the simulation, as opposed to 'random'. 'Narrow' distributions ignore the influence of noise, the latter which is included in 'broad' distributions. SZ-R is used solely for the construction of SZ-1 and SZ-1r, as described in the text. By construction, SZ-1r is equivalent to SZ-ref.}
\label{tab:miscdistros}
\end{table*}

One aim of our work is to test whether the weak lensing mass bias distribution is only sensitive to the radial miscentring distribution as a whole, or whether inherent correlations between the lensing data and SZE or X-ray centers (e.g. regarding the direction of the offset) affect the former. To this end, we randomize the distributions SZ-0 and X-0 by assigning to each target a miscentring from a different, randomly selected, target. The resulting distributions, which we shall call SZ-0r and X-0r, are on average identical to SZ-0 and X-0, while the resulting mass bias distributions may be significantly different. We also construct a randomized distribution, SZ-1r, corresponding to SZ-1.  While SZ-1 does contain a large amount of randomization (namely that coming from SZ-R), we still refer to this distribution as "non-randomized" for simplicity. All miscentring distributions used in this work are summarized in \tabl~\ref{tab:miscdistros}.

In the following subsections, we describe the radial model of reduced shear used to derive masses, and how this model is modified to account for miscentering. 

\subsection{Radial model}
\label{sec:meth:radialmodel}

A direct reconstruction of the projected mass distribution of a cluster of galaxies from weak lensing data (ellipticities) is only possible up to a degeneracy transformation. This so-called mass sheet degeneracy (e.g. \citealt{1988ApJ...327..693G, 1995A&A...294..411S}) is one reason why parametric models are often used when reconstructing masses. A common choice is the azimuthally symmetric Navarro-Frenk-White (NFW) profile \citep{1997ApJ...490..493N}, which has proven to reasonably match both dark matter only \citep{2001MNRAS.321..559B,2012MNRAS.423.3018P,2014ApJ...797...34M,2016MNRAS.457.4340K,2017MNRAS.469.3069G} and hydro-dynamical \citep{2014MNRAS.437.2328B,2016MNRAS.456.3542T} simulations.

Some radial density models provide better agreement with simulated halos near the centers of clusters (e.g. \citealt{2018ApJ...859...55C}). This, however, plays an insignificant role in weak lensing studies \citep{2017MNRAS.465.3361H}, since the central region is usually not included in the analysis. Excising the center also helps in mitigating the effects of inaccurate magnification corrections as well as of miscentring. There are also various alternative models for the outskirts of galaxy clusters (e.g., \citealt{2008arXiv0807.3027T,2014ApJ...789....1D}, and references therein). For reasons of simplicity, we shall make use exclusively of the NFW model in this study. Our conclusions are generally not affected by this, as our general framework of methods can be extended to any azimuthally symmetric model of projected mass density.

The NFW model parameterizes the mass density $\rho(r)$ at physical radius $r$ as 
\begin{equation}
    \rho(r) = \frac{M_\Delta}{4\pi f(c_\Delta)} \frac{1}{r(r+\frac{r_{\Delta}}{c_{\Delta}})^2},
\end{equation}
where $c_{\Delta}$ is the so-called concentration parameter, and 
$f(c_{\Delta}) = \log(1+c_\Delta) - c_\Delta/(1+c_\Delta)$.

Projecting the three-dimensional density model onto the sky plane yields the mass surface density as
\begin{equation}
\label{eq:projectednfw}
    \Sigma(R) = 2 \int_{0}^{\infty} \rho \left( \sqrt{R^2+\zeta^2}\right) \text{d} \zeta,
\end{equation}
where $R$ is now a projected radius and $\zeta$ is in the direction of the line of sight. Analytic expressions for the projected surface density and shear of the NFW profile are provided by \cite{1996A&A...313..697B} and \cite{2000ApJ...534...34W}. 

To determine masses, we use the projected density \eqref{eq:projectednfw} of the NFW model and its associated prediction for reduced tangential shear \eqref{eq:gammafrommass} in radial bins within a range [$\rmin$,~$\rmax$], where the latter can be varied. While we have no uncertainties on the tangential reduced shear (because we have added no noise), we use radial weights based on the projected relative areas of the annuli, and fit the masses using MCMC sampling.   

The concentration parameter $c_{\Delta}$ is often marginalized over in practice, especially at low signal-to-noise ratios. In this work, we use a redshift-dependent concentration$-$mass relation to tie $M_{\Delta}$ to $c_{\Delta}$, enabling a one-parameter fit. From both simulations and observations, only a weak mass dependence of the concentration parameter has been found, and in addition a large scatter \citep{2001MNRAS.321..559B,2008MNRAS.390L..64D,2012MNRAS.423.3018P,2013ApJ...766...32B,2014MNRAS.441.3359D,2014MNRAS.441..378L,2015ApJ...799..108D,2016MNRAS.460.1214L,2017ApJ...840..104S,2019ApJ...871..168D,2019MNRAS.486.4001R}. 

In spite of this weak mass dependence, \citet{2022MNRAS.509.1127S} found that the weak lensing mass bias is strongly dependent upon the concentration$-$mass relation used, in particular, some relations lead to a stronger mass dependence in the bias. In this work, we use the relation of \cite{2015ApJ...799..108D}, with the corrected parameters set of \cite{2019ApJ...871..168D}, noting that while other choices will likely affect the absolute normalization of the mass bias distribution, they will not affect our main results concerning differences in this normalization when comparing different miscentring distributions. 

\subsection{Miscentered models}
\label{sec:meth:miscmodel}

We define the "true" center of a halo, simulated or otherwise, as the bottom of the gravitational potential well of the halo. In simulations, this coincides approximately with the position of the most bound particle\footnote{Because there are finitely many particles in a simulation, the position of the most bound particle will not correspond exactly to the bottom of the potential well. For the simulations used in this work, the difference is negligible.}. In practice, this true center is not known with arbitrary accuracy (other than in simulations), and we have miscentrings on the order of tens to hundreds of kpc, depending on what center proxy is used (the SZE peak, the X-ray peak or centroid, the BCG, the peak of the reconstructed convergence, or other proxies). Here, we derive a model to predict the average tangential shear around a miscentered coordinate.

\begin{figure}
\centering
\includegraphics[width=0.8\columnwidth,clip=True,trim={0 0 0  -20}]{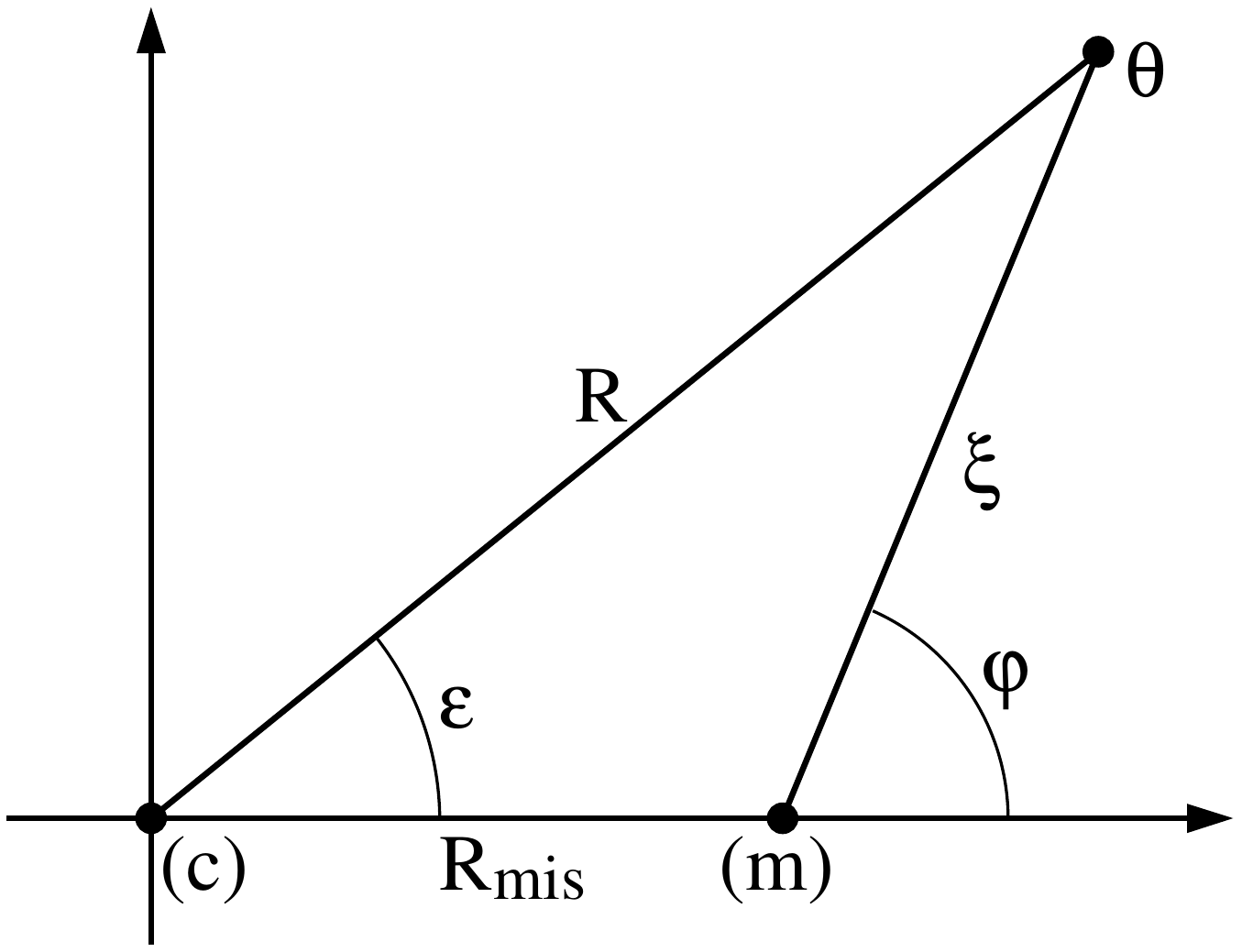}
\caption{\label{fig:rmis} Illustration of the geometry of a miscentered projected azimuthally symmetric mass density profile. The task is to predict the shear (tangential and cross terms), with respect to the miscentered center position $\mathrm{(m)}$ at a coordinate $\theta$ defined by the radius $x$ and the position angle $\varphi$. $\mathrm{(c)}$ marks the true center of a theoretical azimuthally symmetric profile with $g_{\times}=0$ everywhere by definition. The coordinate system is constructed such that the point $\mathrm{(m)}$ is located along the horizontal axis, as only the miscentring radius (not the direction) is of importance in fitting miscentered profiles.} 
\end{figure}

We define a coordinate axis along the displacement of the true center by an amount $\rmis$ as illustrated in Fig.~\ref{fig:rmis}.
The aim is to calculate the tangential and cross shear, $g_{\mathrm{t}}^{\mathrm{(m)}}(\xi,\varphi)$ and $g_{\times}^{\mathrm{(m)}}(\xi,\varphi)$, with respect to the miscentered position $\mathrm{(m)}$, at a coordinate $\theta$ defined by the radial distance $\xi$ and the position angle $\varphi$, and to express these shears in terms of the tangential shear $g_{\mathrm{t}}^{\mathrm{(c)}}$ of the NFW model with respect to the true center $\mathrm{(c)}$ (the cross shear of the centered NFW model is zero everywhere). The position angle $\varepsilon$ of the centered model is measured counterclockwise from the same coordinate axis, at the position of the true center. 

Combining \eqns~{(\ref{eq:shearxt}) and (\ref{eq:shear12})} with the fact that $g_{\times}^{\mathrm{(c)}}=0$ leads to 
\begin{align}
    \label{eq:gmt_from_gc}
    g_{\mathrm{t}}^{\mathrm{(m)}}(\xi)  &=  g_{\mathrm{t}}^{\mathrm{(c)}}(R) \left[ \cos{(2 \varepsilon)} \cos{(2 \varphi)} + \sin{(2 \varepsilon)} \sin{(2 \varphi)} \right]; \\
    \label{eq:gmx_from_gc}
    g_{\times}^{\mathrm{(m)}}(\xi) &=  g_{\mathrm{t}}^{\mathrm{(c)}}(R) \left[ \sin{(2 \varepsilon)} \cos{(2 \varphi)} - \cos{(2 \varepsilon)} \sin{(2 \varphi)} \right],
\end{align}
\noindent where 
\begin{equation}
    R^2 = \rmis^2 + \xi^2 + 2 \rmis \xi \cos{\varphi}.
\end{equation}
Using the laws of sines and cosines, we shall additionally have (see Fig.~\ref{fig:rmis})
\begin{align}
\label{sincos}
\xi^2 &= \rmis^2 + R^2 - 2 \, \rmis \, R \, \cos{\varepsilon}; \\ 
\sin{\varepsilon} &= \frac{\xi \, \sin{\varphi}}{R},
\end{align}
whereby (\ref{eq:gmt_from_gc}) and \eqref{eq:gmx_from_gc} can be expressed as
\begin{equation}
    g_{\mathrm{t}}^{\mathrm{(m)}}(\xi,\varphi) = g_{\mathrm{t}}^{\mathrm{(c)}}(R)\left(  1 - \frac{2 \, \rmis^2 \, \sin^2{\varphi}}{R^2} \right)
\end{equation}
and
\begin{equation}
    g_{\times}^{\mathrm{(m)}}(\xi,\varphi) = - g_{\mathrm{t}}^{\mathrm{(c)}}(R) \frac{2 \, \rmis \sin{\varphi} \left(\rmis \cos \varphi + \xi \right)}{R^2},
\end{equation}
respectively. 

In practice, the miscentring radius $\rmis$ is not known for any individual target. In this work, for the purpose of fitting masses, we take $\rmis$ to be the mean radius of the miscentring distribution (\citealt{2021MNRAS.507.5671G}), the latter which is known to us (but for which we also never have perfect knowledge of in practice). 

 Predicting the average tangential reduced shear at radius $\xi \ne R_{\rm{mis}}$ (with respect to the miscentered position) in a thin annulus amounts to evaluating the integral
\begin{equation}
g_{\mathrm{t},\rm{avg}}^{\mathrm{(m)}}(\xi) = \frac{1}{2 \pi}  \int_{\varphi=0}^{2 \pi} g_{\mathrm{t}}^{\mathrm{(m)}}(\xi,\varphi) \, {\rm{d}} \varphi = \frac{1}{\pi} \int_{\varphi=0}^{\pi} g_{\mathrm{t}}^{\mathrm{(m)}}(\xi,\varphi) \, {\rm{d}} \varphi,
\end{equation}
where the second equality is due to $g_{t}^{(m)}(\xi,\varphi)$ being an even function of $\varphi$ at constant $\xi$. For $\xi = \rmis$, we must formally take the limit
\begin{equation}
   g_{\mathrm{t},\rm{avg}}^{\mathrm{(m)}}(\xi) = \lim_{\phi \xrightarrow{} 0} \frac{1}{\pi} \int_{\varphi=\phi}^{\pi} g_{\mathrm{t}}^{\mathrm{(m)}}(\xi,\varphi) \, {\rm{d}} \varphi.
\end{equation}

\begin{figure}
\centering
\includegraphics[width=0.99\columnwidth,clip=True,trim={0 0 0 0}]{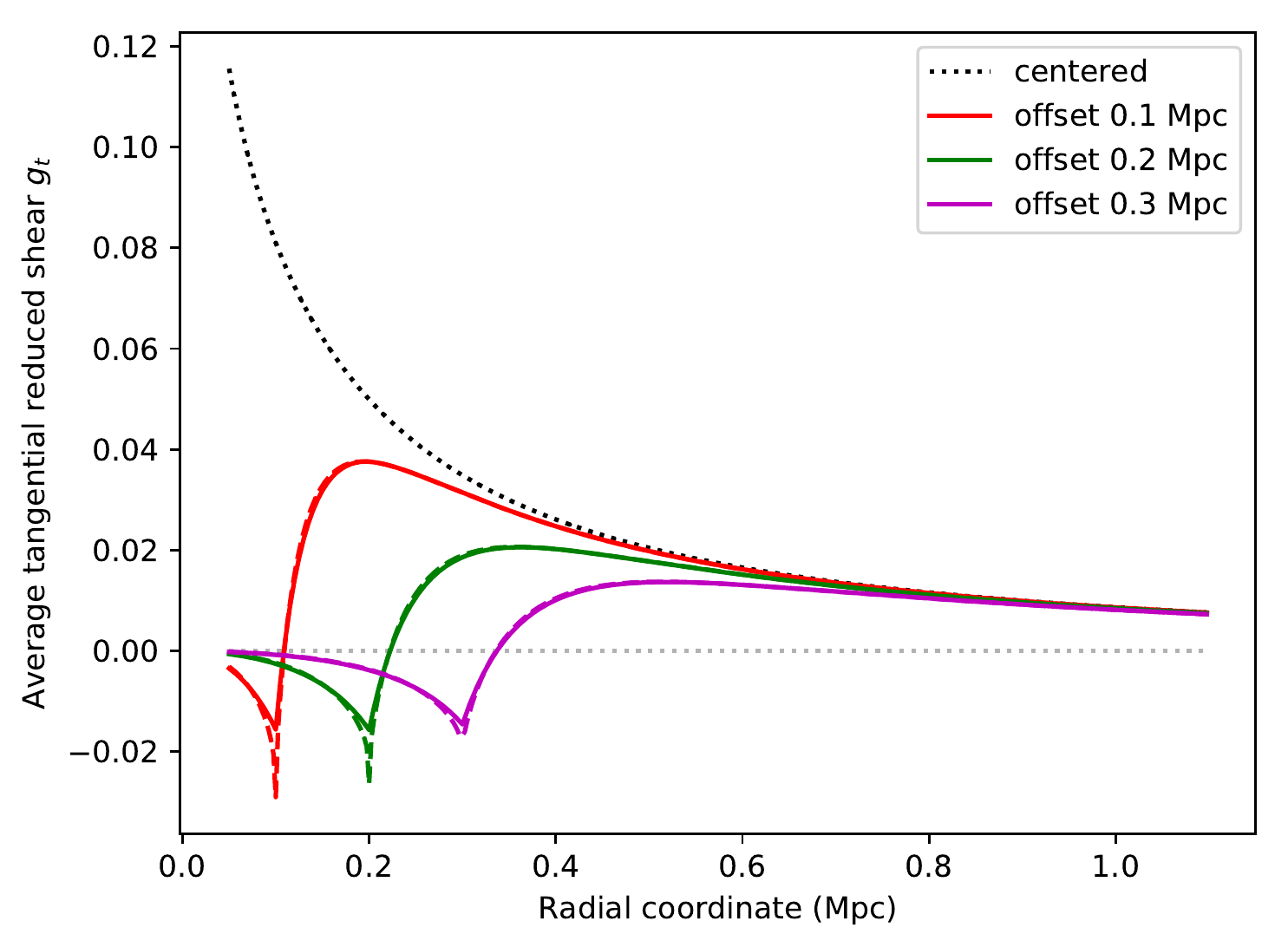}
\caption{\label{fig:g_avg_differentrmis} Miscentered reduced shear profiles for an ideal NFW halo with mass $M_{500} = 0.7 \times 10^{14} \msun$ and concentration $c=3.8$, placed at $z=0.7$ with $\beta = 0.3$, for different miscentring radii. Solid lines: NFW approximation with constant convergence inside the miscentring radius. Dashed lines: full NFW model. The black dotted line represents a corresponding profile without miscentring.} 
\end{figure}

Although we compute the reduced shear only outside some chosen radius $r_{\rm{min}}$ around the miscentered coordinate, we can come very close to the center of the properly centered profile, potentially leading to computational problems as the reduced shear diverges. Following \citet{2021MNRAS.507.5671G}, we alleviate this problem by setting the convergence $\kappa$ at radial values (of the centered profile) less than the miscentring radius to the value at that radius. The effect of this  is small, as can be seen in \fig\thinspace\ref{fig:g_avg_differentrmis}.

\subsection{Mass bias modeling}
\label{sec:meth:massbias}

We define the weak lensing mass bias as the ratio
\begin{equation}
    \bias = \frac{\wlmass}{\truemass},
\end{equation}
where $\wlmass$ and $\truemass$ are the measured and true masses, respectively, for the same value of the over-density $\Delta$. For a sample of targets, $\bias$ can be viewed as a random variable. In the remainder of this work, such random variables are assumed to be either normal or log-normal distributed. Since masses of galaxy clusters are typically not negative, a log-normal distribution seems intuitively more sound, and also makes it easier to account for the mass bias when fitting mass-observable relations to data. 

A framework for quantifying the consistency of the empirical mass bias distributions resulting from simulations with normal and log-normal distributions is outlined in \sect~\ref{sec:meth:wasserstein}.

To avoid confusion, we use the symbols ($\munormal$,~$\sigmanormal$) to respectively signify estimated means and standard deviations in linear space, and similarly ($\mulognormal$,~$\sigmalognormal$) in logarithmic space.

\subsection{Over-correction of mass}
\label{sec:meth:overcorr}

Given an estimate of the mass bias distribution, it is a relatively simple exercise to correct a measured mass, propagating the uncertainties of the measurements and taking the shape of the bias distribution into account (effectively broadening the mass uncertainty). Using random miscentring distributions (SZ-0r, SZ-1r and X-0r) may yield corrections different from those derived based on corresponding non-randomized miscentring distributions (SZ-0, SZ-1 and X-0). In this subsection, we are concerned with quantifying such differences, taking corrections from non-randomized miscentrings as the reference. Mass over-correction, therefore, is taken as the (positive) relative change in estimated mass when using randomized miscentring distributions.

We consider only the estimated means of the mass bias distributions, and estimate the amount by which a mass would on average be over-corrected (using the mean mass bias) due to a random miscentring distribution being used. In the case of a Gaussian model of the mass bias distribution, we define the over-correction as 
\begin{equation}
    \overcorrectionnormal = \frac{\munormal^{\nonrandommiscentring}}{\munormal^{\randommiscentring}} - 1,
\end{equation}
while for a log-normal model of the mass bias distribution the corresponding over-correction factor is
\begin{equation}
    \overcorrectionlognormal = \exp(\mulognormal^{\nonrandommiscentring} - \mulognormal^{\randommiscentring})-1.
\end{equation}
Here, the superscripts $\nonrandommiscentring$ and $\randommiscentring$ denote non-random and randomized miscentrings, respectively. Perhaps contrary to intuition, these expressions are not the same, since $\mulognormal \neq \log(\munormal)$.

\subsection{Quantifying the mass bias distributions}
\label{sec:meth:wasserstein}

The mass bias distributions resulting from our analysis are purely empirical, and it is not strictly necessary to use any parametric model to quantify them. However, for purposes such as fitting mass-observable relations, it is more practical to have closed-form expressions. In this work, we quantify the observed mass bias  in terms of normal and log-normal distributions, and test how well they indeed conform to the latter. To this end, we make use of the so-called Wasserstein $p$-distance (e.g. \citealt{villani2008}). 

For an observed (sampled) mass bias distribution $u(x)$ and a theoretical distribution $v(x)$, 
the Wasserstein distance in one dimension ($p=1$) can be written as
\begin{equation}
\label{eq:wasserstein}
    l_1(u,v) = \int_{-\infty}^{\infty} |U(x)-V(x)| \dd{x},
\end{equation}
where $U(x)$ and $V(x)$ are the cumulative distribution functions (CDF) corresponding to $u$ and $v$, and $x$ is the independent variable (in our case, the mass bias). Intuitively, this can be understood as a cost function, in the sense that $l_1$ quantifies how far the samples of $u$ must to me moved in order for the result to conform to $v$ (e.g. \citealt{2015arXiv150902237R}). In practice, we always set $v$ to be a normal distribution. To test for a log-normal distribution, we take $\log(x)$ as the independent variable.

To define the theoretical distribution $v$, we must define the mean $\mu$ and variance $\sigma^2$. Because it is not known a priori whether $u$ was actually drawn from a normal distribution, we estimate $(\mu,\sigma)$ by minimizing $l_1(u,v)$. This is not an unbiased estimator, as discussed further below. 

While a low value of the Wasserstein distance $l_1$ indicates a high degree of consistency with the theoretical distribution, it is not straightforward to quantify the level of consistency, as the empirical distribution has a finite number of samples. In particular, one would expect $l_1$ to scale inversely with the square root of the number of samples $n$, and to increase linearly with $\sigma$. To quantify this effect, we first estimate the Wasserstein distance in the case where the empirical $u$ is actually drawn from the theoretical $v$. 

Given $\sigma$ and $n$, we thus draw many random instances $\hat{u}$, each with $n$ samples, from $v$, and evaluate $l_1$ for each. The resulting distribution of $l_1$ can now be compared to a Wasserstein distance measured from a $u$ of unknown origin. We find that $\log(l_1)$ follows a normal distribution 
\begin{equation}
    \label{eq:wasserstein_is_lognormal}
    \log (l_1) \sim \mathcal{N}(\mu_l(\sigma,n), \sigma_l^2(\sigma,n)),
\end{equation}
enabling us to quantify the level of agreement with $u$ in terms of its deviation $\bar{w}$ from $\mu_l$ in units of $\sigma_l$ as
\begin{equation}
    \label{eq:wasserstein_w}
    \bar{w}(u,v) = \frac{l_1(u,v) - \mu_l}{\sigma_l}
\end{equation}
(equivalently, one can express this as the probability that $u$ was drawn from $v$, using the CDF of the normal distribution). A large value of $\bar{w}$ would indicate a poor agreement, while a low negative value would indicate that the data may be corrupted in some way. 

Because we are minimizing the Wasserstein distance to derive the parameters of $u$, the expectancy value of $\bar{w}$ will be artificially low. We test for this bias by again drawing many instances of $\hat{u}$ from $v$, now minimizing the Wasserstein distance $l_1$ to estimate $\mu$ and $\sigma$, arriving in each realization at a somewhat different theoretical distribution $\hat{v}$. Proceeding to compute $\bar{w}(\hat{u},\hat{v})$ in a large number of instances yields an approximation of the distribution of $\bar{w}$. We find that $\bar{w}$ is biased low by a constant $\sim 1.4$, independently of $\sigma$ and $n$. We thus take 
\begin{equation}
    \label{eq:wasserstein_wcorr}
    w(u,v) = \bar{w}(u,v) + 1.4 
\end{equation}
as our measure of consistency between $u$ and $v$. In our results, we refer to $\wnormal$ and $\wlognormal$ in quantifying the consistency with normal distributions of $\bias$ and $\logbias$, respectively.

\section{Results}
\label{sec:results}

We begin this section by investigating whether there is a significant mass dependence in the weak lensing mass bias distribution, focusing on the narrow miscentring distributions SZ-0(r)\footnote{We use the shorthand notation SZ-0(r) to signify that we consider both SZ-0 and SZ-0r. Similarly, SZ(X) means that we consider both X and SZ miscentring distributions.} and X-0(r) as well as on perfectly centered halos (we refer to \tabl~\ref{tab:miscdistros} for a description of the miscentring distributions). After determining the differences in bias going from SZ(X)-0 to the randomized SZ(X)-0r, we proceed to add noise and consider differences in bias using the SZ-1 and SZ-1r distributions. We also investigate the level of (log-)normality of the bias distributions in the sense of \sect~\ref{sec:meth:wasserstein}, using both centered and miscentered mass fitting. In the aforementioned analysis, we use a default fitting range from $\rmin=0.5$ Mpc to $\rmax=1.5$ Mpc (physical), assuming a uniform distribution of background galaxies and no shape or shear noise, and a lensing efficiency $\beta=0.3$. Finally, we vary the fitting range ($\rmin$ and $\rmax$) to investigate if and to what extent this alleviates the difference in mass bias when using randomized offsets compared to the ones determined from the ICM of the simulation. 

\subsection{Mass dependence in the weak lensing mass bias}
\label{sec:results:massdep}

\begin{figure}
\centering
\includegraphics[width=0.97\columnwidth,clip=True,trim={5 0 0 0}]{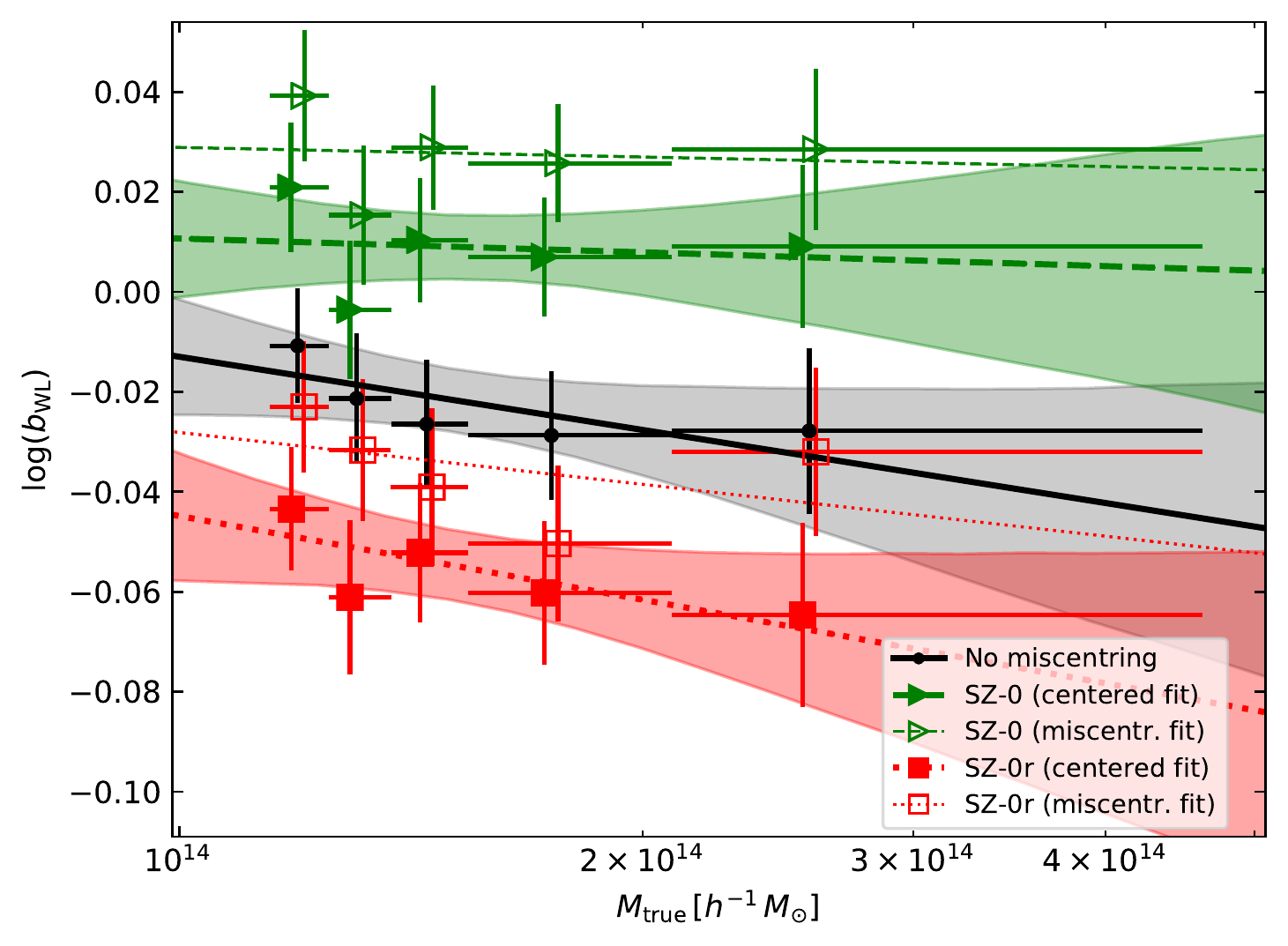} \\
\includegraphics[width=0.97\columnwidth,clip=True,trim={5 0 0 0}]{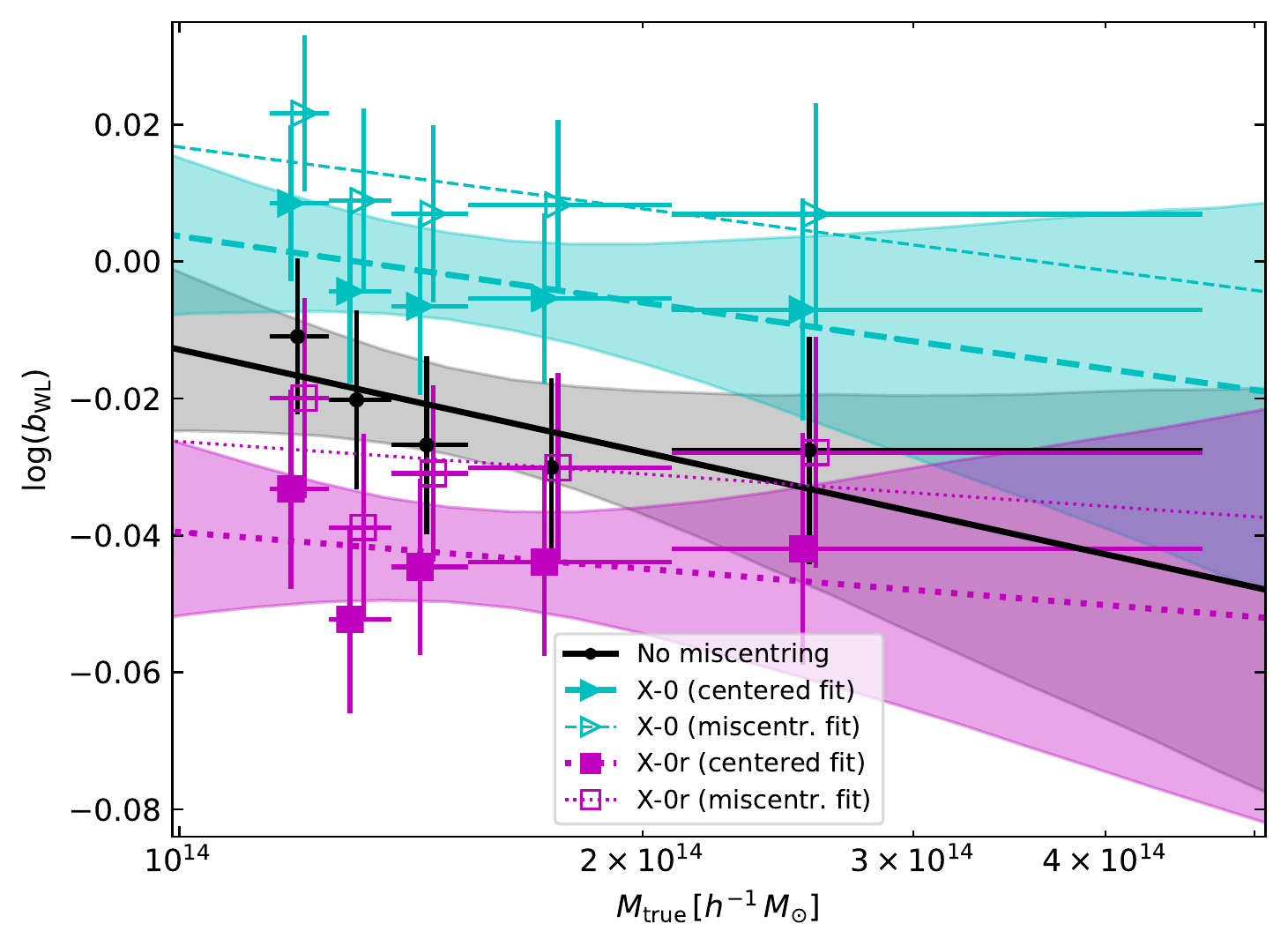}
\caption{\label{fig:massdep0} Mass dependence of the weak lensing mass bias, as determined using the SZ-0(r) (top panel) and X-0(r) (bottom panel) miscentring distributions. The black points and the solid lines represent the bias in the absence of miscentring (same in both panels). Triangles and dashed lines indicate the mass bias for non-randomized miscentering distributions (SZ-0 and X-0), while squares and dotted lines correspond to the randomized miscentring distributions SZ-0r and X-0r. Filled (open) symbols and thick (thin) lines correspond to centered (miscentered) fits. To make the plots more intelligible, confidence regions are shown only for centered fits. All masses are $\mfh$.}
\end{figure}

\begin{table*}
\centering
\begin{tabular}{l c r@{$\pm$}l cc c r@{$\pm$}l cc} 
 \hline
             &  \multicolumn{5}{c}{centered mass fit}  & \multicolumn{5}{c}{miscentered mass fit}  \\
\cmidrule(lr){2-6} \cmidrule(lr){7-11}
 Miscentring &  slope      & \multicolumn{2}{c}{$\mulognormal$ at median $\mfh$} & $\wlognormal$ & $\wlognormal^{\mathrm{corr}}$ & slope      & \multicolumn{2}{c}{$\mulognormal$ at median $\mfh$} & $\wlognormal$ & $\wlognormal^{\mathrm{corr}}$ \\
 \hline
 None    & $-0.022 \pm 0.024$ & $-0.021$ & $0.006$ & 4.27 & 4.42 & $-$ & \multicolumn{2}{c}{$-$} & $-$ &  $-$ \\
 \hline
 SZ-0    & $-0.004 \pm 0.023$ & $+0.009$ & $0.006$ & 4.23 & 4.25 & $-0.003 \pm 0.023$ & $+0.028$ & $0.006$ & 4.31 & 4.33 \\
 SZ-0r   & $-0.024 \pm 0.026$ & $-0.054$ & $0.007$ & 2.89 & 3.01 & $-0.015 \pm 0.025$ & $-0.034$ & $0.007$ & 3.57 & 3.59 \\
        &                    & $\overcorrectionlognormal = \mathbf{+0.065}$ &  $\mathbf{0.009}$ & & &          & $\overcorrectionlognormal = \mathbf{+0.064}$ & $\mathbf{0.009}$ & & \\
 \hline
 X-0     & $-0.014 \pm 0.023$ & $-0.001$ & $0.006$ & 4.01 & 4.08 & $-0.013 \pm 0.023$ & $+0.012$ & $0.006$ & 4.03 & 4.10 \\
 X-0r    & $-0.007 \pm 0.024$ & $-0.042$ & $0.007$ & 3.36 & 3.40 & $-0.007 \pm 0.025$ & $-0.029$ & $0.007$ & 3.38 & 3.41 \\
 &      & $\overcorrectionlognormal = \mathbf{+0.042}$ & $\mathbf{0.009}$ & & &          & $\overcorrectionlognormal = \mathbf{+0.042}$ & $\mathbf{0.009}$ & &\\
 \hline
\end{tabular}
\caption{\label{tab:massdep0} Mass dependence of the mean mass bias, in the sense of \eqn~\eqref{eq:loglogmassdep}, for narrow (noiseless) miscentring distributions. The logarithm of the mass bias is given at the median true $\mfh$. Values shown in boldface represent the fractional overestimation of masses, after bias correction, when using randomized as compared to non-randomized miscentring distributions (the two entries immediately above each boldface value in the table). The Wasserstein distances $\wlognormal$ and $\wlognormal^{\mathrm{corr}}$ pertain to log-normal distributions before and after correcting for the fitted mass dependence, respectively.}
\end{table*}

A mass dependence in the WL mass bias would lead to a misrepresentation of the corresponding bias distribution if not accounted for. In this subsection, we investigate whether there is such a mass dependence. Using the default setup and the miscentring distributions SZ-0(r) and X-0(r), in addition to a setup without any miscentring, we fit for a mass dependence of the form 
\begin{equation}
    \logbias = A + B ~ \log{\truemass},
    \label{eq:loglogmassdep}
\end{equation}
using all pairs of true masses ($\mfh$) and bias values determined from the NFW mass fitting, using linear regression with equal weights. This assumes the distribution of the logarithmic bias at constant mass has no skewness -- in effect assume a log-normal distribution, coupled with a mass dependence modeled by a power law. Additionally, for visualization, we divide the determined individual bias measurements into five bins, such that the bins contain approximately equal numbers of targets. 

The results are shown in \fig\thinspace\ref{fig:massdep0} and \tabl~\ref{tab:massdep0}. From the determined slopes, we see that for all miscentring distributions used here, the results are consistent with no mass dependence. While the best-fit slopes are all negative, this can be explained by the data not being independent -- regardless of the miscentring distribution, the targets are the same in each run. 

In the absence of miscentring, the normalization of the mass bias distribution is close to zero (masses are biased low by approximately 2\% on average). This is not a general result, but a specific one pertaining to our particular setup including radial mass fitting range, redshift and choice of concentration$-$mass relation. 

With random miscentring, $\bias$ decreases (masses are biased low to a greater extent) compared to when the gravitational centers are assumed to be known with arbitrary accuracy. Interestingly, with non-random miscentring distributions, the bias increases to levels higher then in the absence of miscentring. In particular, comparing random to non-random miscentrings, we find a difference of around 6\% in the SZE case (SZ-0 vs. SZ-0r) and a difference of about 4\% in the X-ray case (X-0 vs. X-0r), regardless of the type of mass fit (centered vs. miscentered). We also note that using miscentered fits yields higher values of $\bias$, as we would expect. 

In \tabl~\ref{tab:massdep0} we also report the normalization of the fitted mass dependence as the value of $\logbias$ at median $\mfh$. To see whether the removal of the tentative mass dependence improves log-normality in the mass bias distribution, we use the inverse of \eqn~\eqref{eq:loglogmassdep} and compute the Wasserstein measure with and without this correction ($\wlognormal^{\mathrm{corr}}$ and $\wlognormal$, respectively). In no case does the mass bias distribution conform better to a log-normal distribution after correcting for the  determined mass dependence. In the rest of this work, we shall assume that the mass bias, in the limited range of masses considered here, is independent of mass.

\subsection{Narrow miscentring distributions}
\label{sec:results:narrowmisc}

In the previous subsection, the normalization of the WL mass bias was been determined using linear regression applied to the logarithmic bias as a function of true mass. We now proceed to compute this normalization by minimizing the Wasserstein distance, as described in \sect~\ref{sec:meth:wasserstein}, first for the narrow miscentring distributions (this subsection), and subsequently in \sect~\ref{sec:results:broadmisc} for the broader miscentring distributions SZ-1 and SZ-1r, assuming that any mass dependence is negligible. 

\begin{figure*}
\centering
\includegraphics[width=1.00\columnwidth,clip=True,trim={0 7 5 0}]{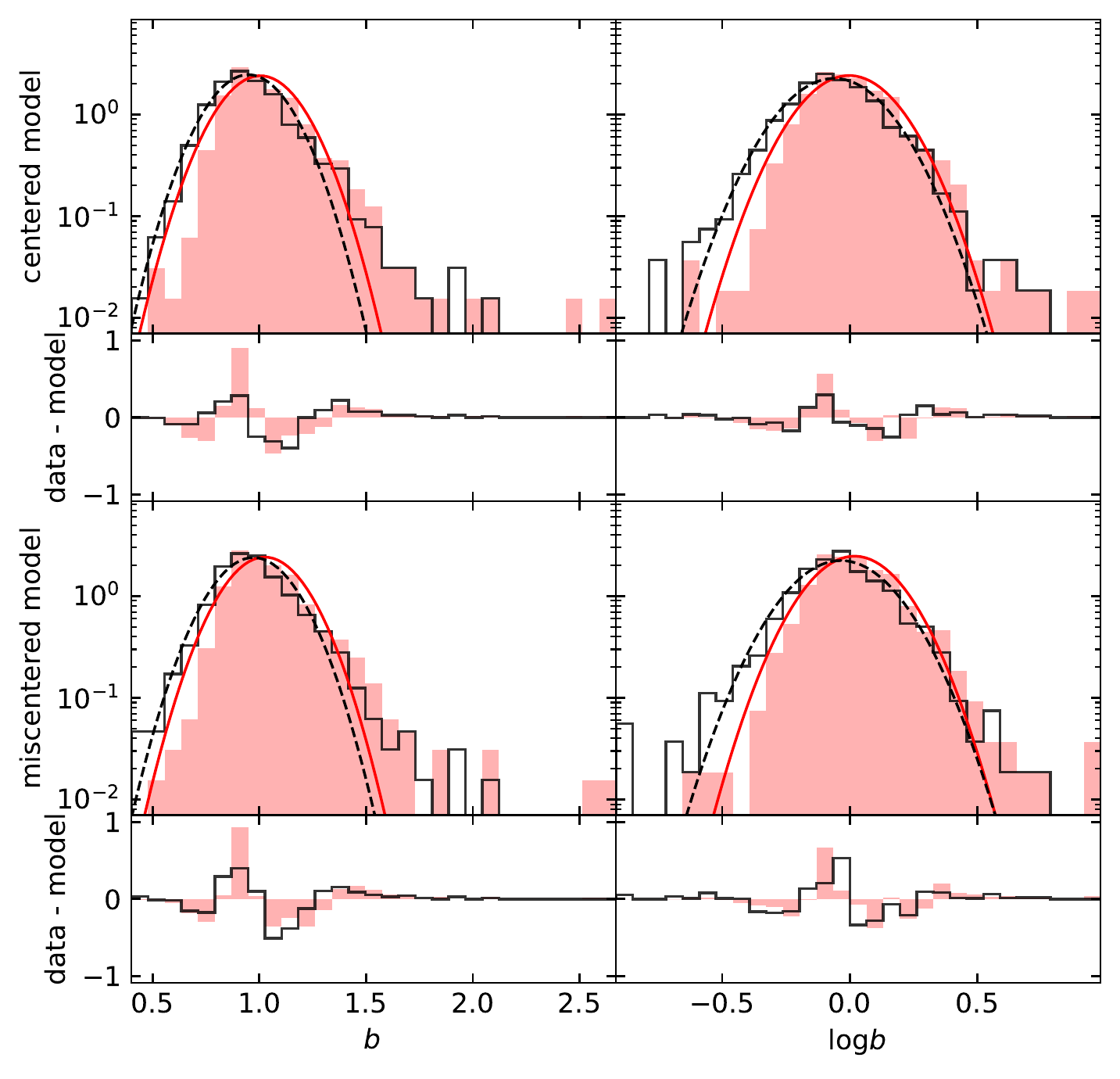} 
\includegraphics[width=1.00\columnwidth,clip=True,trim={0 7 5 0}]{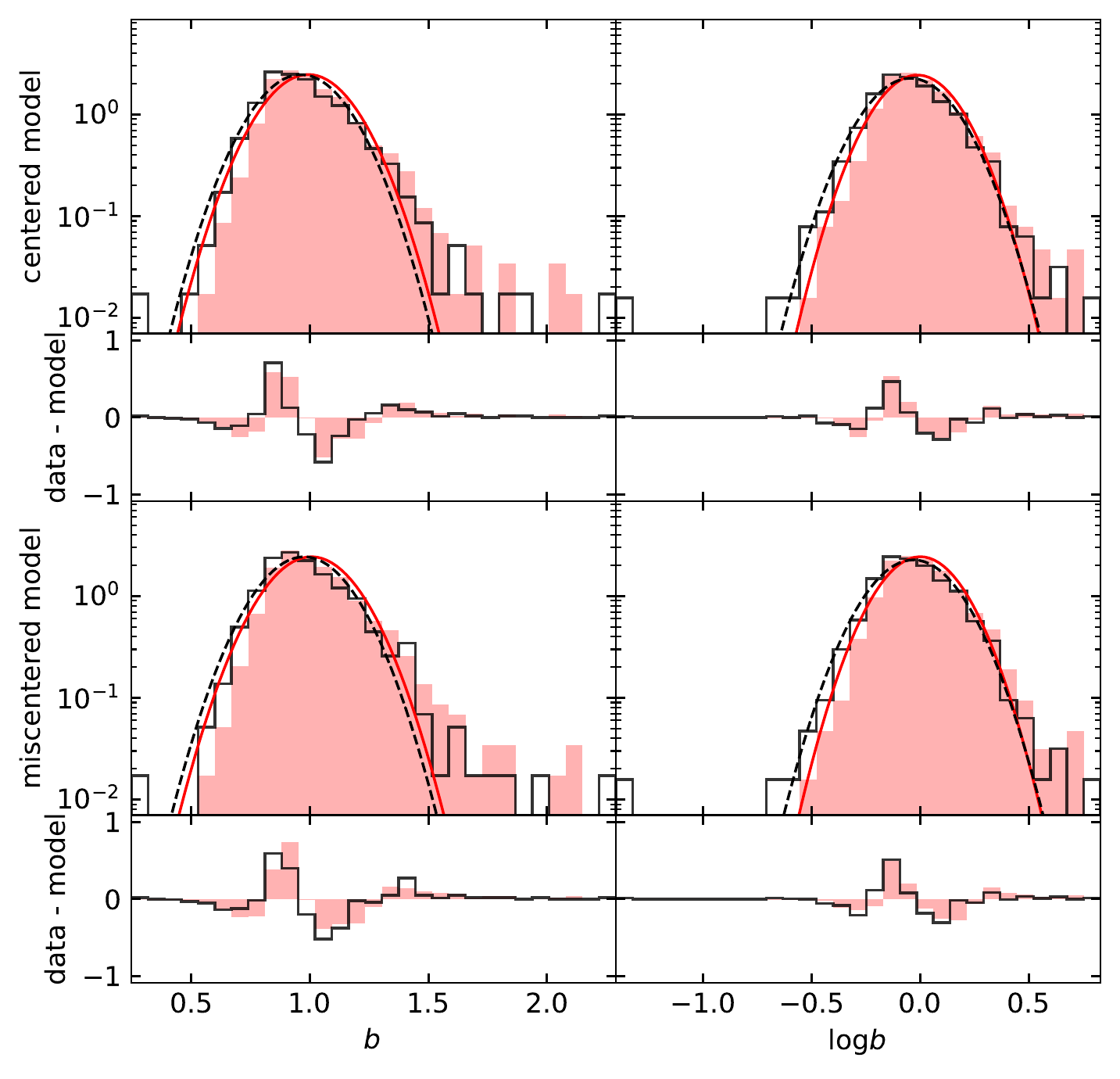} \\
\caption{\label{fig:biashists_narrow} \textit{Left panel:} weak lensing bias distributions derived from simulations, for the two narrow SZE miscentring distributions SZ-0 (non-randomized; filled red histograms) and SZ-0r (randomized; black step histograms). For the top and bottom rows, centered and miscentered models were respectively used in the mass fitting. The red filled curves (corresponding to SZ-0) and the black dashed curves (corresponding to SZ-0r) show the best-fit (in the sense of minimizing $w$) normal (left sub-panels) and log-normal (right sub-panels) distributions. Differences between the observed and theoretical distributions are shown in linear scale in the smaller sub-panels below each histogram. \textit{Right panel:} Corresponding bias distributions derived with the narrow X-ray miscentring distributions X-0 (filled histograms, red solid curves) and X-0r (black step histograms, dashed black curves).}  
\end{figure*}

Thus far, we have seen that random miscentring tends to yield mass bias distributions in which $\munormal$ and $\mulognormal$ are underestimated, leading to overestimated masses after correction. In  \fig\thinspace\ref{fig:biashists_narrow}, we show the full weak lensing mass bias distributions resulting from using the narrow miscentring distributions SZ-0(r) and X-0(r) in linear and log-space of $\bias$, for miscentered and centered mass fitting. We also show best-fit normal distributions, derived from minimizing $\wnormal$ and $\wlognormal$. The corresponding  results are given in the first part of  \tabl~\ref{tab:bdistros_with_wasserst}, where we also give the mass over-correction factors $\overcorrectionnormal$ and $\overcorrectionlognormal$. We summarize these results as follows. 

\begin{enumerate}
\item The bias distributions derived for the SZE and X-ray based centers are very similar in terms of normalization and scatter. This is not surprising, as the miscentring distributions have similar shapes and widths.
\item In general, log-normal distributions provide better fits than normal distributions. However, this is in part due to our choice of $\rmin$ and $\rmax$, as discussed further in \sect~\ref{sec:results:rdep}. Applying no miscentring at all results in a very low value of $\wlognormal$, and a high value of $\wnormal$. 
\item An over-correction of mass ($\overcorrectionnormal$ and $\overcorrectionlognormal$) is present in all combinations of randomized and non-randomized miscentring distributions used. The over-correction factor $\tau$ is consistent within the SZE and X-ray miscentrings, regardless of whether $\bias$ or $\logbias$ is considered. There is, however, a significant difference in that the SZE based miscentrings consistently yield higher values of $\tau$.
\item Comparing miscentered to centered mass fitting, there is no significant difference in the results, other than the absolute normalization of the mass bias distribution.
\end{enumerate}

\subsection{Broad miscentring distributions}
\label{sec:results:broadmisc}

In this subsection, we investigate the effects on the mass bias of using the broad miscentring distributions SZ-1(r). It may seem reasonable to suspect that the extra randomness induced by various noise terms, such as instrumental effects, sky noise and primary CMB anisotropies in the case of centers derived from the SZE, would alleviate the effects of taking a random miscentring distribution as opposed to one correlating tightly with the SZE signal of the simulation. 

\begin{figure}
\centering
\includegraphics[width=1.00\columnwidth,clip=True,trim={0 7 5 0}]{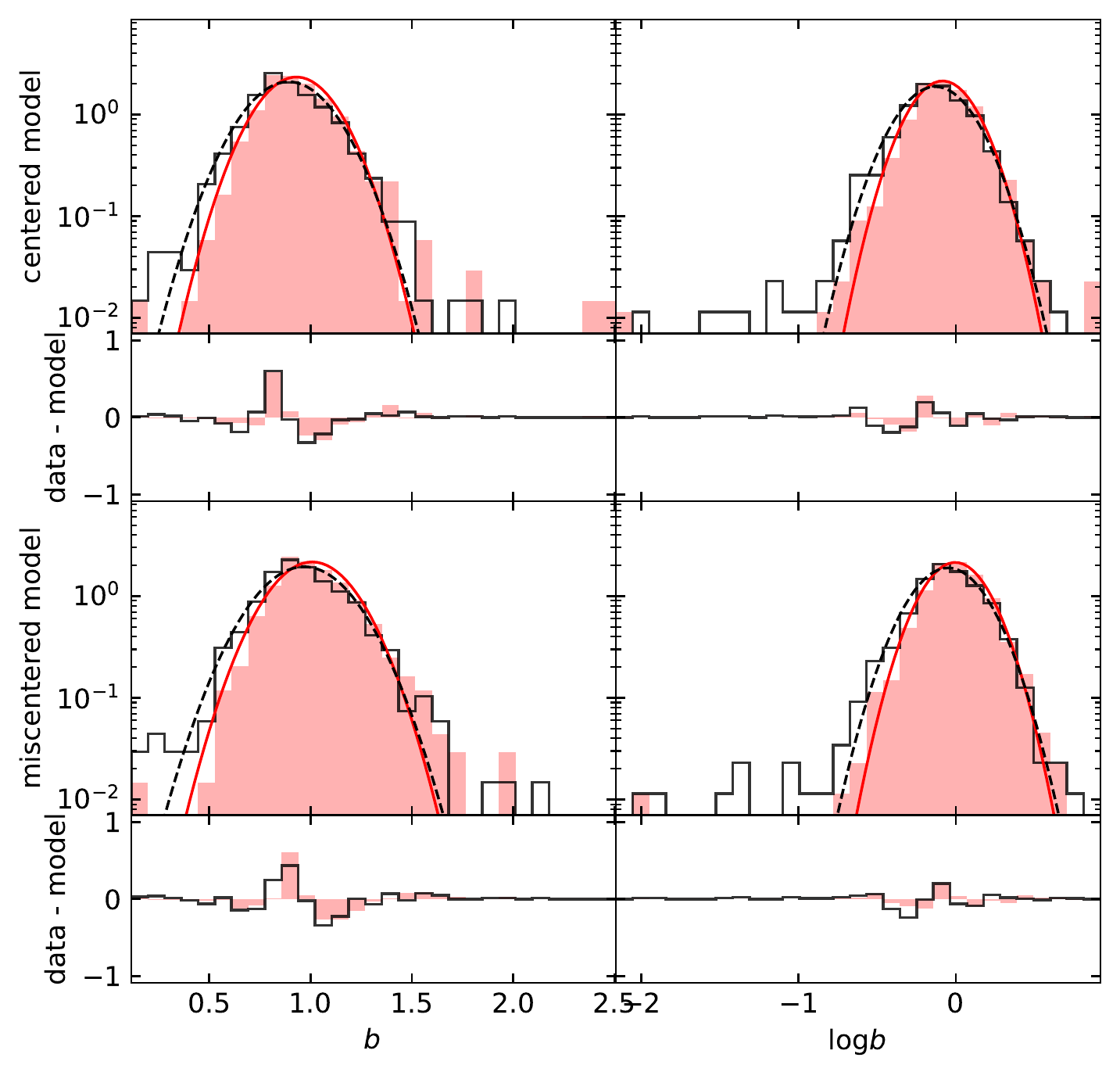} \\
\caption{\label{fig:biashist_SZ_broad} As \fig\thinspace\ref{fig:biashists_narrow}, but for the broad SZE miscentring distributions SZ-1 (filled histograms, red solid curves) and SZ-1r (black step histograms, dashed black curves).}  
\end{figure}

We show the results of using broad miscentring in the second part of \tabl~\ref{tab:bdistros_with_wasserst} and in \fig\thinspace\ref{fig:biashist_SZ_broad}. The over-correction of mass is reduced slightly, but still significant, ranging from 4.2 to 5.7 percent for our default choice of the radial fitting range. The mass bias distribution still conforms better to a log-normal than a normal distribution when using non-randomized miscentring, while for randomized miscentring there is hardly any difference. As with the narrow miscentring distributions, there is very little difference using centered vs. miscentered fitting, other than in the normalization of $\bias$. 

\begin{table*}
\centering
\begin{tabular}{l r  r@{$\pm$}l r@{$\pm$}l c  r@{$\pm$}l r@{$\pm$}l c} 
 \hline
&  & \multicolumn{5}{c}{normal distribution} &  \multicolumn{5}{c}{log-normal distribution} \\
\cmidrule(lr){3-7} \cmidrule(lr){8-12}
Miscentring   & Fit & \multicolumn{2}{c}{$\munormal$} & \multicolumn{2}{c}{$\sigmanormal$} & $\wnormal$ & \multicolumn{2}{c}{$\mulognormal$} & \multicolumn{2}{c}{$\sigmalognormal$} & $\wlognormal$ \\
\hline
None  & $\cfit$  & 0.978&0.005 &  0.155&0.004 &   6.12 & $-$0.030&0.006 &  0.161&0.004 &   2.32 \\
\hline
\hline
X-0   & $\cfit$  & 0.993&0.006 &  0.161&0.004 &   5.67 & $-$0.013&0.006 &  0.163&0.004 &   4.08 \\
X-0r  & $\cfit$  & 0.963&0.006 &  0.162&0.004 &   5.39 & $-$0.045&0.006 &  0.175&0.004 &   3.38 \\ 
 &  &  $\overcorrectionnormal=~$+\textbf{0.031}&\textbf{0.009} & 
\multicolumn{2}{c}{} & & $\overcorrectionlognormal=~$+\textbf{0.033}&\textbf{0.009} & \multicolumn{2}{c}{} & \\
\hline
X-0   & $\mfit$  & 1.007&0.006 &  0.163&0.004 &   5.67 &    0.001&0.006 &  0.163&0.004 &   4.03 \\
X-0r  & $\mfit$  & 0.976&0.006 &  0.164&0.004 &   5.41 & $-$0.031&0.006 &  0.175&0.004 &   3.44 \\
 &  & $\overcorrectionnormal=~$+\textbf{0.032}&\textbf{0.009} & \multicolumn{2}{c}{} & & $\overcorrectionlognormal=~$+\textbf{0.033}&\textbf{0.009} & \multicolumn{2}{c}{} & \\
\hline
SZ-0  & $\cfit$  &  1.005&0.006 &  0.166&0.004 &   5.98 & $-$0.002&0.006 &  0.165&0.004 &   4.14 \\
SZ-0r & $\cfit$  &  0.948&0.006 &  0.162&0.004 &   5.15 & $-$0.061&0.006 &  0.176&0.004 &   2.91 \\ 
 &  & $\overcorrectionnormal=~$+\textbf{0.060}&\textbf{0.009} & \multicolumn{2}{c}{} & & $\overcorrectionlognormal=~$+\textbf{0.061}&\textbf{0.009} & \multicolumn{2}{c}{} & \\
 \hline
SZ-0  & $\mfit$  & 1.026&0.006 &  0.165&0.004 &   5.92 &    0.019&0.006 &  0.161&0.004 &   4.29 \\
SZ-0r & $\mfit$  & 0.972&0.006 &  0.166&0.004 &   5.16 & $-$0.035&0.006 &  0.178&0.004 &   3.44 \\
 &  &  $\overcorrectionnormal=~$+\textbf{0.056}&\textbf{0.009} & \multicolumn{2}{c}{} & & $\overcorrectionlognormal=~$+\textbf{0.056}&\textbf{0.009} & \multicolumn{2}{c}{} & \\
\hline
\hline
SZ-1  & $\cfit$  &   0.930&0.006 &  0.171&0.004 &   4.52 & $-$0.084&0.006 &  0.186&0.005 &   2.41 \\
SZ-1r & $\cfit$  &   0.892&0.007 &  0.190&0.005 &   3.57 & $-$0.130&0.007 &  0.212&0.005 &   3.76 \\ 
 &  &  $\overcorrectionnormal=~$+\textbf{0.043}&\textbf{0.007} & \multicolumn{2}{c}{} & & $\overcorrectionlognormal=~$+\textbf{0.057}&\textbf{0.010} & \multicolumn{2}{c}{} & \\
 \hline
SZ-1  & $\mfit$  &  1.008&0.006 &  0.184&0.005 &   4.67 & $-$0.004&0.006 &  0.186&0.005 &   2.45 \\
SZ-1r & $\mfit$  &  0.968&0.007 &  0.205&0.005 &   3.56 & $-$0.049&0.007 &  0.210&0.005 &   3.60 \\
 &   & $\overcorrectionnormal=~$+\textbf{0.042}&\textbf{0.010} & \multicolumn{2}{c}{} & & $\overcorrectionlognormal=~$+\textbf{0.046}&\textbf{0.010} & \multicolumn{2}{c}{} & \\
\hline
\end{tabular}
\caption{\label{tab:bdistros_with_wasserst} Results of fitting normal and log-normal distributions, in the sense of minimizing $w$, for all miscentring distributions considered in this work, with $(\rmin,~ \rmax)=(0.5,1.5)$ Mpc. Pairs of non-randomized and randomized miscentring distributions are between horizontal lines. As in \tabl~\ref{tab:massdep0}, each boldface value indicates the fractional mass over-correction $\overcorrection$ for the receding pair of miscentring distributions. In the second column, $\cfit$ and $\mfit$ refer to centered (\sect~\ref{sec:meth:radialmodel}) and miscentered (\sect~\ref{sec:meth:miscmodel}) mass fits, respectively.}
\end{table*}

\subsection{Dependence on radial range}
\label{sec:results:rdep}

The effects of miscentring are strongest at small radii, and tend to get averaged out at larger radii as the difference between the true center and the chosen center become less important. It is thus reasonable to assume that the differences in mass bias distributions, both from using centered vs. miscentered simulations as well as from using randomized vs. non-randomized miscentring distributions, will become less pronounced as we increase one or both of the inner and outer fitting radii $\rmin$ and $\rmax$. Here, we explore the effects of varying these two parameters in the ranges $0.2~\rm{Mpc} \leq \rmin \leq 1.0~\rm{Mpc}$ and $1.0~\rm{Mpc} \leq \rmax \leq 2.2~\rm{Mpc}$, with the requirement that $\rmax - \rmin \geq 0.4~\rm{Mpc}$. We carry out mass fits exclusively with the miscentered model described in \sect~\ref{sec:meth:miscmodel}, and consider only the broad miscentring distributions SZ-1 and its randomized counterpart SZ-1r.

While excising a large chunk of the cluster center is desirable in terms of lowering the systematic uncertainty due to miscentring, it is of course desirable to use as much of the lensing data as possible to maximize the signal-to-noise ratio (SNR) to lower the statistical uncertainty of the mass. As we are using a uniform distribution of background galaxies, weights of radial bins will increase linearly, while the reduced shear simultaneously decreases. The net effect is that the contribution to the signal-to-noise ratio of each radial bin is a somewhat complicated function of the density profile. Rather than modeling this dependence, we use the weights and signals to estimate how the ensamble average uncertainty in mass varies with $\rmin$ and $\rmax$.

\begin{figure}
\centering
\includegraphics[width=1.00\columnwidth,clip=True,trim={5 5 5 0}]{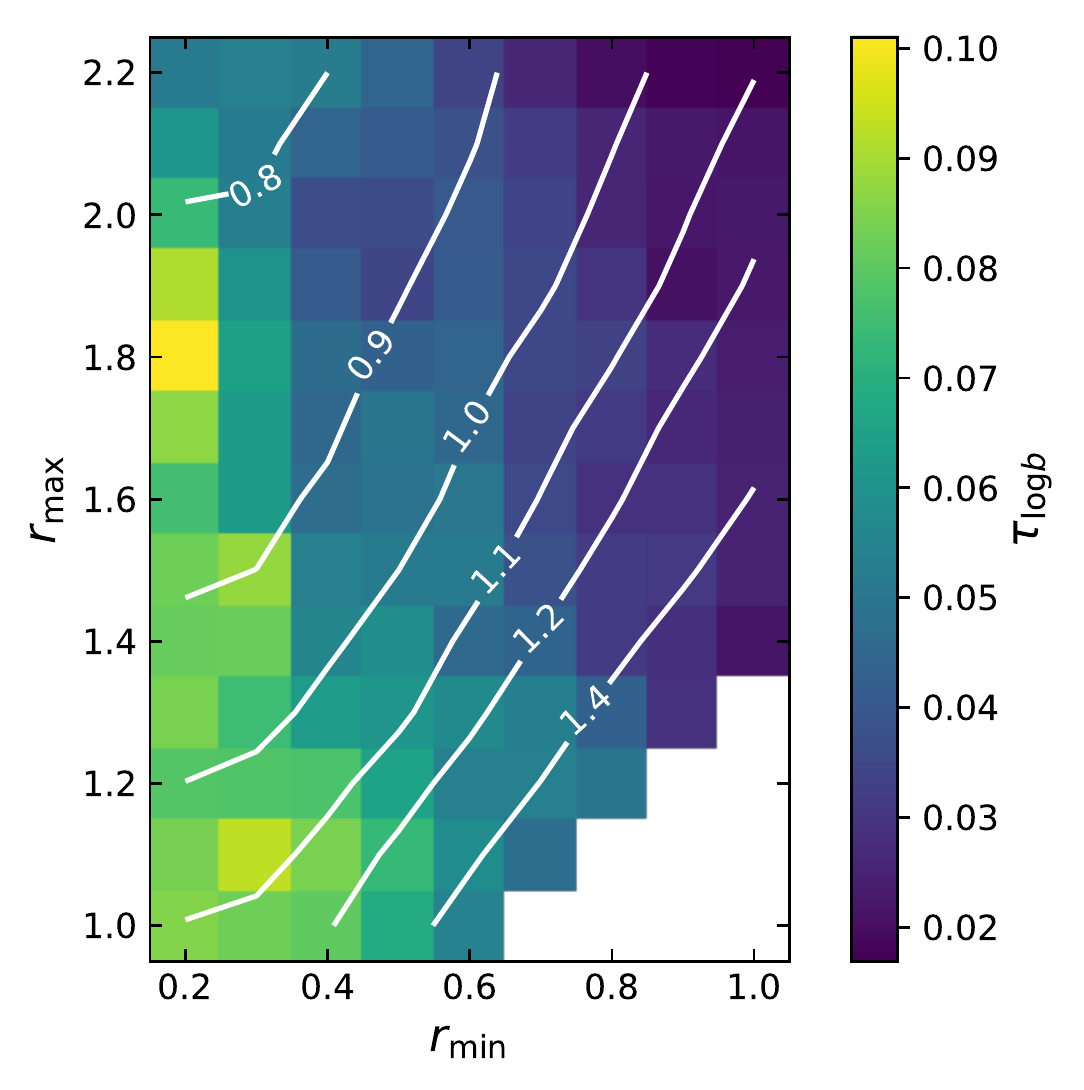} \\
\caption{\label{fig:tau_rminrmax} Over-correction parameter $\overcorrectionlognormal$ (color) in logarithmic space as a function of the radial mass fit range determined by $\rmin$ and $\rmax$ (Mpc physical), with miscentring from the distributions SZ-1 (non-random) and SZ-1r (randomized). The contours show the relative uncertainty on mass for the different fit ranges, normalized to one at $(\rmin,\rmax)=(0.5,1.5)$ Mpc. Masses were fitted using the miscentered model.}  
\end{figure}

As expected, the mass over-correction due to the randomized miscentring distribution steeply decreases with increasing $\rmin$, and also to a smaller extent with increasing $\rmax$ (\fig\thinspace\ref{fig:tau_rminrmax}). With $\rmin \gtrsim$ 0.8 Mpc, however, the choice of $\rmax$ seems less important, as is apparent in \fig\thinspace\ref{fig:tau_rminrmax}. When considering the largest values of the two radii allowed by our constraints, the over-correction is around 2$\%$. Using the radial weights and average signal in the radial bins of reduced shear of the simulated halos in our sample, we can estimate the relative uncertainty in mass ensuing from the different choices of the two radii. We normalize this mass uncertainty at our default setup with $\rmin = 0.5$ Mpc and $\rmax = 1.5$ Mpc. From \fig\thinspace\ref{fig:tau_rminrmax}, we can see that choosing the maximal values of the radii increases the mass uncertainty by around 40\% when compared to maximizing $\rmax$ while taking $\rmin$ at its lowest value of 0.2 Mpc.

\begin{figure*}
\centering
\includegraphics[height=9.4cm,clip=True,trim={5 5 85 0}]{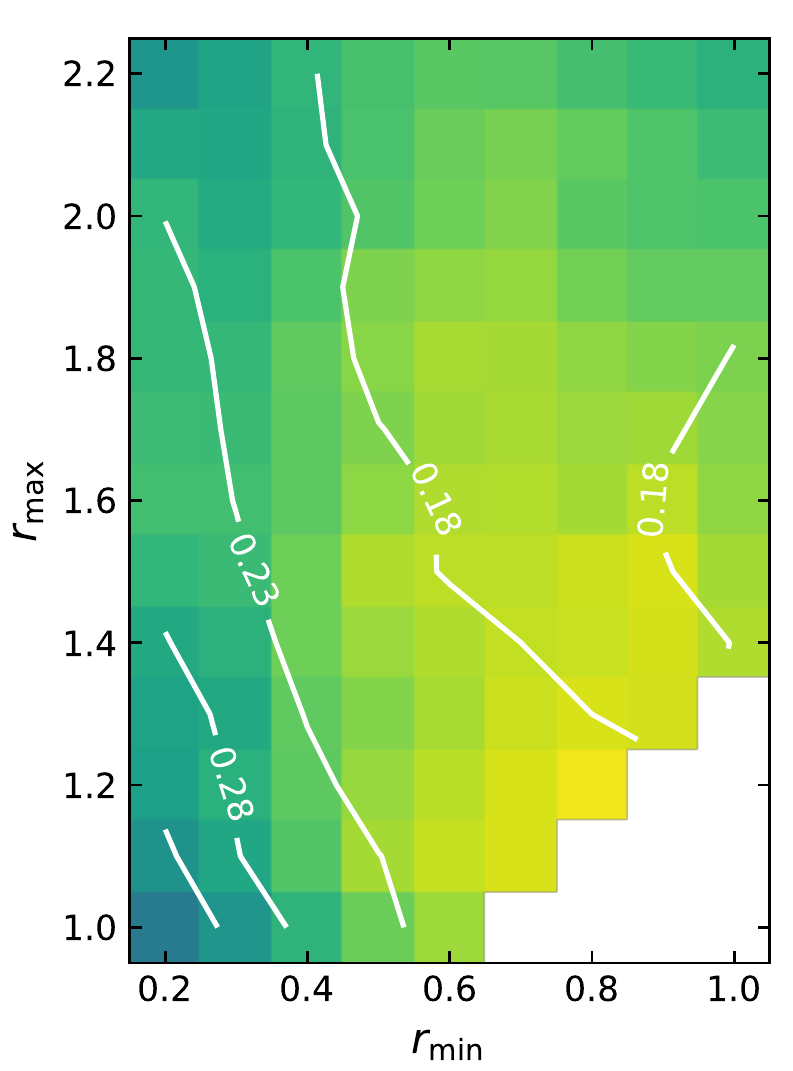}
\hspace{1.5mm}
\includegraphics[height=9.4cm,clip=True,trim={41 5 5 0}]{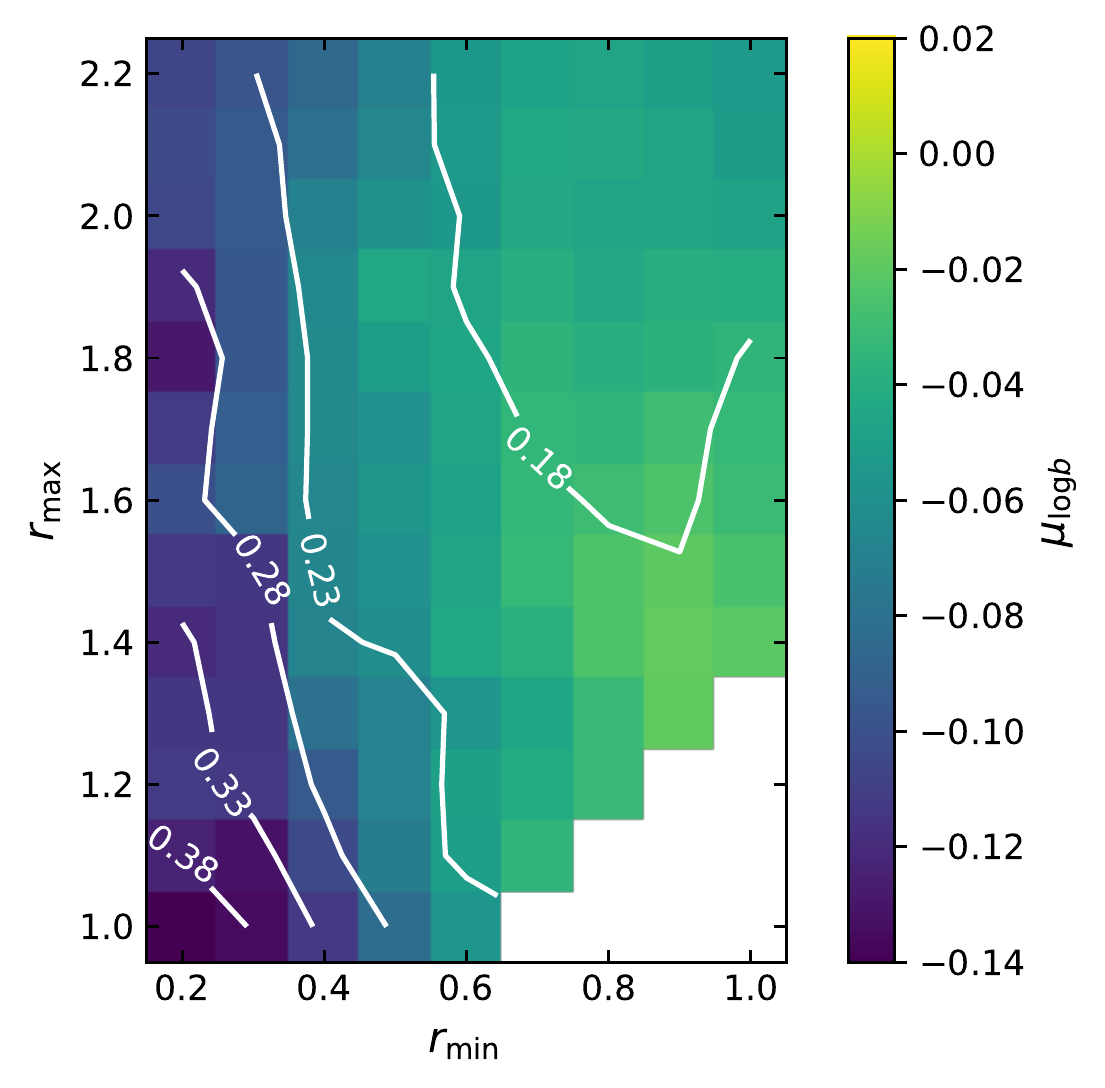}\\
\caption{\label{fig:musigma_rminrmax} Estimated mean $\mulognormal$ (colors) and standard deviation $\sigmalognormal$ (contours) of a log-normal mass bias distribution, as a function of the radial mass fit range ($\rmin$ and $\rmax$, in physical Mpc). Two miscentring distributions were used here; the non-randomized broad SZE miscentring distribution SZ-1 (left panel), and the randomized counterpart SZ-1r (right panel).}  
\end{figure*}

The estimated values of $\mulognormal$, underlying \fig\thinspace\ref{fig:tau_rminrmax}, are shown in color in \fig\thinspace\ref{fig:musigma_rminrmax}, along with the scatter $\sigmalognormal$ as contours. For low values of $\rmin$ and $\rmax$, this scatter is grossly overestimated when using the randomized miscentring distribution SZ-1r, a fact that is not captured by the measure $\overcorrectionlognormal$.

\begin{figure*}
\centering
\includegraphics[height=9.5cm,clip=True,trim={0 5 55 0}]{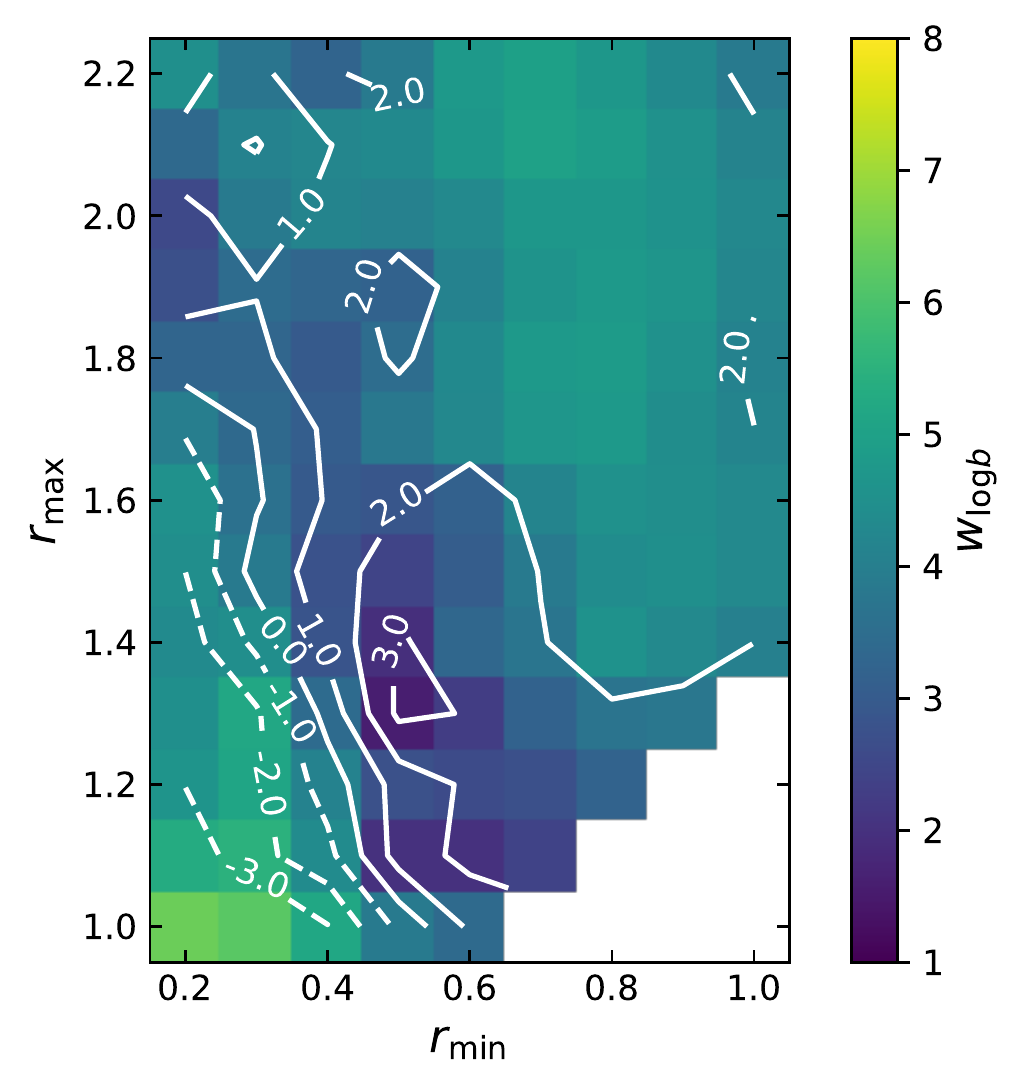}
\includegraphics[height=9.5cm,clip=True,trim={41 5 -10 0}]{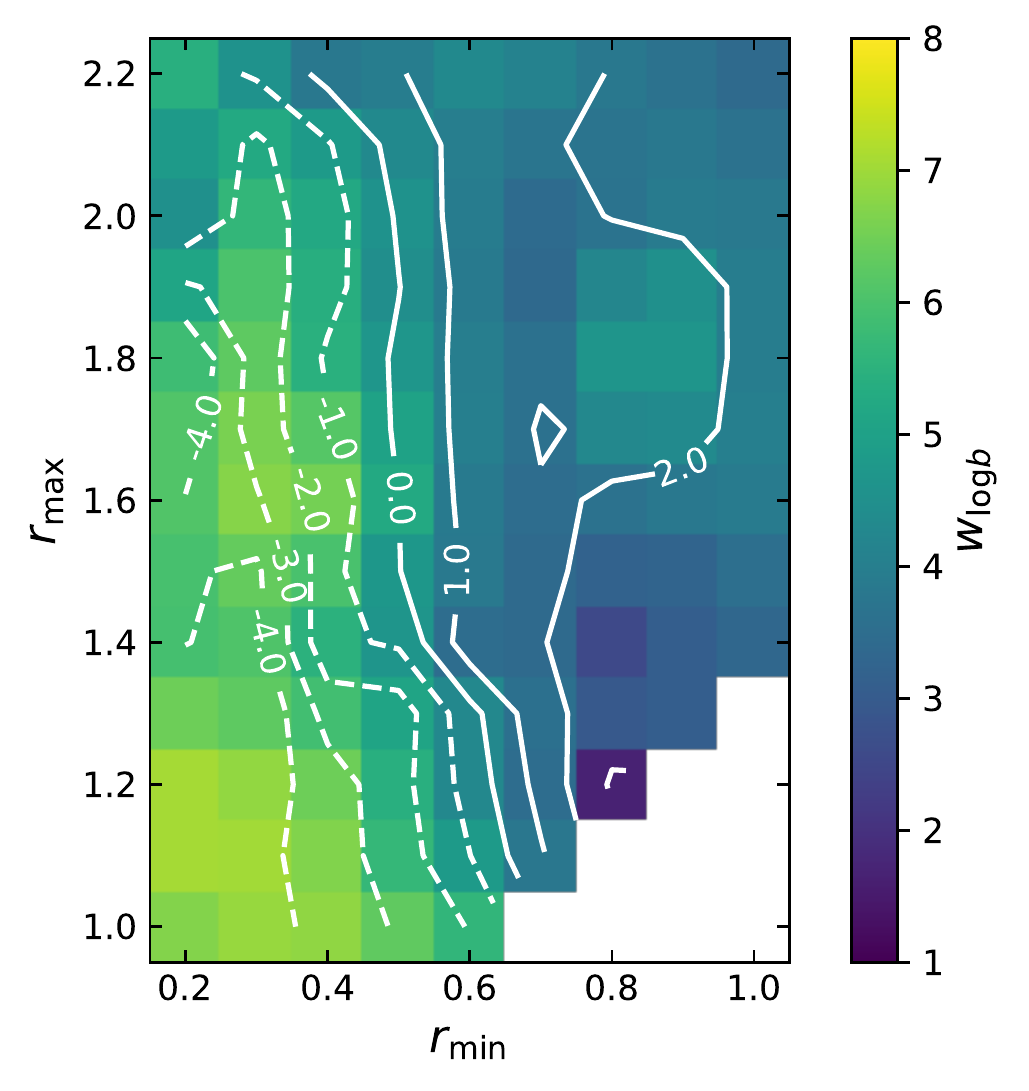}\\
\caption{\label{fig:w_rminrmax} Wasserstein distance based parameter $\wlognormal$, quantifying the number of standard deviations separating the observed mass bias distribution from a log-normal distribution, plotted as a function of the radial mass fit range. A number close to one indicates that the empirical distribution is consistent with a log-normal. To see how consistent the distributions are with Gaussians, we show the difference $\Delta w = \wnormal - \wlognormal$ as contours: a negative value means a Gaussian is a better fit, while a positive value suggests that a log-normal is more suitable. As in \fig\thinspace\ref{fig:musigma_rminrmax}, the left and right panels correspond to the miscentring distributions SZ-1 and SZ-1r, respectively, and $\rmin$ and $\rmax$ are given in physical Mpc.}
\end{figure*}

\fig\thinspace\ref{fig:w_rminrmax} shows the Wasserstein measure $\wlognormal$, the number of standard deviations in excess of being consistent with a log-normal distribution (\sect~\ref{sec:meth:wasserstein}) in color, with overlaid contours indicating the difference $\Delta w = \wnormal - \wlognormal$, to determine whether there are ranges of ($\rmin,~\rmax$) where a normal distribution provides better consistency than a log-normal. The latter is in fact the case for a large part of the parameter space considered; in particular with randomized miscentring, the bias distribution looks more normal than log-normal with a lower radial bound less than approximately $0.5$ Mpc (with a weak dependence on $\rmax$). Interestingly, increasing $\rmax$ does not in general correspond to a decrease in $\wlognormal$, a statement that holds true for both SZ-1 and SZ-1r.   

\section{Discussion}
\label{sect:discussion}

We begin this section by discussing the mass dependence and the shape of the mass bias distribution under different assumptions. Subsequently, we discuss the implications of the over-correction of masses arising from taking randomized miscentring distributions, the latter which has has been the norm in recent studies. We also discuss in some detail the limitations of this work.

\subsection{Mass dependence}
\label{sec:discussion:massdep}

While we have found no significant mass dependence in the limited range of masses considered (due to its limited volume, the Magneticum simulation has few very massive halos), \cite{2018MNRAS.479..890L}, \cite{2021MNRAS.507.5671G} and \cite{2022MNRAS.509.1127S}, found that the mean mass bias does change with mass. The latter results, however, are not directly comparable with our findings as different concentration-mass relations were used (\citealt{2018MNRAS.479..890L} marginalized over the concentration parameter rather than tying it to the mass). In addition, dependencies on other factors, such as the redshift and the radial range of the reduced shear used in the fitting, must be taken into account. For the purpose of this work, we showed that the mass dependence is not strong enough to affect the shape of the bias distribution in a significant way, allowing us to study the shape and normalization of the latter to sub-percent level accuracy within the confines of the simulation and our setup. 

\subsection{Shape of the mass bias distribution}
\label{sec:discussion:shape}

While it has been shown with some confidence from simulations that the weak lensing mass bias distribution closely resembles a log-normal distribution for \textit{perfectly centered} halos \citep{2021MNRAS.507.5671G,2022MNRAS.509.1127S}, the situation seems somewhat more complicated in the presence of miscentring. 
The results of the present work paint a rather complicated picture of how closely the distribution can be represented by a log-normal distribution, with strong dependencies on the choice of the radial range for fitting. Additionally, we would also expect possible dependencies on mass, redshift and the chosen concentration-mass relation. While such dependencies have not been explicitly investigated here, we can state with some confidence that the mass bias distribution is sufficiently complicated that it must be tailored individually to any real observation of a sample of galaxy clusters (including the selection function). To model the shape of the distribution more accurately, one may of course resort to more complicated models, such as the sum of two or more theoretical distribution functions (notwithstanding the fact that log-normal distributions are very appealing when dealing with inherently positive quantities such as mass). Regardless of the model chosen, the first two moments of the distribution can be modeled to high accuracy using hundreds of simulated clusters, and the more pressing issue is thus the discrepancy arising when using randomized miscentring distributions. 

\subsection{Over-correction of masses}
\label{sec:discussion:overcorr}

Given an estimation of the WL mass bias distribution, masses are typically not corrected; instead the inherent mass bias, as determined from simulations or otherwise, is included when correlating weak lensing masses to other observables. Nevertheless, the effect of the latter is exactly the same, namely, to overestimate weak-lensing masses because the correlation between the ICM distribution and the weak lensing shear has not been taken into account. 

This is a bias that has, to our knowledge, not been considered in previous publications. Considering that typical values of $\rmin$ range from $0.3$ to $0.7$ Mpc (e.g. \citealt{2021MNRAS.505.3923S, 2022A&A...668A..18Z, 2023MNRAS.tmp..949C}), a mass over-correction of several percent would be expected, taking our results at face value. As a result, miscentring corrected masses inferred from weak lensing have been overestimated in the past. This has implications for resulting scaling relations and cosmological constraints. We expect, for instance, that a stronger disagreement in the $\sigma_8$ constraints between \textit{Planck} and SPT would have been found by \citet{2019ApJ...878...55B}, had their weak lensing mass bias estimates been corrected for this effect. 

Our results suggest that the correlation between weak lensing, chiefly due to dark matter, and the distribution of the ICM in galaxy clusters are correlated to such a degree that the blind use of miscentring distributions around the gravitational center (the bottom of the gravitational potential well) is not warranted, even in the presence of a robust understanding of such miscentring distributions. \citet{2018MNRAS.474.2635S} speculated that this may be the case, arguing that in the case or merging systems, ICM centers may be located between the mass centers and therefor not in a random offset direction from the deepest point of the gravitational potential. 

While the approach of using miscentring distributions can be robust in general, the present results indicate that random miscentring should not be scattered around the gravitational center (in simulations), but perhaps rather on the center of mass in the ICM (when the latter is used as a center proxy). 

\subsection{Limitations of this work}
\label{sec:discssion:limitations}

We summarize the limitations of this work in the following. 

\begin{enumerate}
\item Our results have been computed for specific setups and can therefore not be used to infer accurate bias corrections for other scenarios. Rather, they indicate the need to take into account the effects of using random miscentring distributions when accounting for the impact of ICM based center proxies on weak lensing mass biases.  
\item We have exclusively considered simulated galaxy clusters at $z=0.67$. It is likely that the impact of random miscentring on the WL mass bias will vary with redshift. Additionally, we have not considered a realistic mass selection of clusters. 
\item Our results, being based upon the use of one hydrodynamic simulation, can at most be as good as the latter. Comparing results from different simulations would be a fist step in determining the robustness of our claims.  
\item Related to the previous point, the miscentring distributions used here are themselves not well constrained. In fact, a major part of the error budget relating to WL mass bias comes from the uncertainty in such distributions (e.g. \citealt{2019ApJ...878...55B,2021MNRAS.505.3923S,2022A&A...668A..18Z}).
\item We have not accounted for realistic distributions of background galaxies in the sky plane, but instead assumed a uniform distribution of the latter. This limitation is of course easy to overcome by examining a set of simulations tailored to the observations at hand. 
\item Our analysis is limited to noiseless radial profiles of reduced shear, allowing a sub-percent level of accuracy in the determination of the parameters of the WL mass bias distribution. While \cite{2022MNRAS.509.1127S} showed that this approach is valid under the assumption of a log-normal bias distribution, it is not known whether a different distribution may be noise dependent. \cite{2018MNRAS.479..890L} found some evidence of a noise dependence, albeit without inspecting the bias distribution directly (the latter which is challenging in the presence of noise), and by using a different approach to the projected mass density profile (marginalizing over the NFW concentration parameter rather than tying it to the mass as we do here).     
\item Due to the fact that our narrow X-ray miscentring distribution is broader (at low radii) than the corresponding reference distribution, we have not considered a broad X-ray miscentring distribution. It seems unlikely, however, judging from the impact of random miscentring in the broad SZE miscentring distribution on the WL mass bias, that random X-ray centers would lower the corresponding WL mass bias to a sub-percent level.
\item Our maximum choice of $\rmax$ is quite low compared to some ground-based surveys and future space based surveys. In spite of the fact that $\rmin$ seems to have more impact in absolute terms (changing it by a value $\Delta r$ vs. changing $\rmax$ by the same amount), the inner radius can of course be increased further than what we have considered if more data are available in the cluster outskirts. 
\end{enumerate}

\section{Summary and Conclusions}
\label{sec:conclusion}

We summarize our methods and main results as follows:

\begin{enumerate}

\item We use the Magneticum Pathfinder hydrodynamic simulations at redshift $z=0.67$ to study the effects of miscentring on the distribution of biases in weak lensing cluster mass estimates. In particular, we investigate the effect of using random miscentring distributions, as opposed to miscentring offsets derived directly from the ICM component of the simulations on a cluster to cluster basis, the latter which preserves correlations between different centers. Our sample consists of 275 halos, projected along three orthogonal axes, yielding a final sample of 825 quasi-independent targets, allowing constraints better than $1\%$ on the normalization of the mass bias distribution.  

\item Azimuthally symmetric models are used throughout, although we construct a miscentered model that can be fit to weak lensing data assuming a known mean miscentring radius (but with no assumed knowledge of the direction of any individual offset). We find no significant differences in our results when comparing centered and miscentered models, except in the normalization of the mass bias distribution.

\item We set up a rudimentary framework, based on the Wasserstein 1-distance between probability distributions, to test how well the resulting mass bias distributions conform to normal or log-normal distributions. We find that neither model can reliably be used universally, but deviations from log-normality are generally small when a sufficiently large portion of the cluster center is excised from the data. The latter also mitigates the effects of miscentring, both in the sense of leading to smaller differences when using random miscentring distributions vs. individual offsets, and in the sense that a mass bias distribution derived in the absence of miscentring becomes a more accurate approximation. 

\item We find that randomly assigning offsets to simulate a miscentring, causes weak lensing masses to be overestimated by 2--6 \%, depending on the radial range of the weak lensing data (reduced shear) used in fitting masses. This may have severe implications for large-scale surveys of clusters of galaxies, and suggests that the miscentring problem is not yet understood to such a degree that sub-percent accuracy is possible. In particular, we have used one slice of one simulation here, while another simulation may give a slightly different answer to the question of mass bias. 

\end{enumerate}

Accurately calibrating masses of dark matter halos has become a matter of increasing concern in the previous two decades, as statistical uncertainties continue to be reduced due to larger space- and ground-based surveys with increasing sensitivity.  

Our results, while pertaining specifically to biases arising from ignoring the directional correlation between the ICM and the lensing shear, may possibly go to show something more general, namely that major sources of systematic bias on the order of several percent can be overlooked, potentially leading to discrepancies for which it may be an option to search for missing systematics before asking more fundamental questions about our understanding of the Universe (the latter which, of course, should always be put into serious question). 

\section*{Acknowledgements}

We would like to thank Klaus Dolag and his collaborators on the Magneticum Pathfinder simulations, in particular Alex Saro and Veronica Biffi, for useful hints on the use of the simulations. We thank Peter Schneider for discussions and detailed comments on the manuscript. 

MS and TS acknowledge support from the German Federal Ministry for Economic Affairs and Energy (BMWi) provided
through DLR under projects 50OR2002, 50OR2106, 50OR2302,
and 50QE2002 as well as support provided by the
Deutsche Forschungsgemeinschaft (DFG, German Research Foundation) under grant 415537506. 

TS also acknowledges support provided by the Austrian Research Promotion Agency (FFG) and the Federal Ministry of the Republic of Austria for Climate Action, Environment, Mobility, Innovation and Technology (BMK) via the Austrian Space Applications Programme with grant numbers 899537 and 900565.  

%%%%%%%%%%%%%%%%%%%%%%%%%%%%%%%%%%%%%%%%%%%%%%%%%%

\section*{Data availability}

The data underlying this work will be shared on reasonable request to the corresponding author.

%%%%%%%%%%%%%%%%%%%% REFERENCES %%%%%%%%%%%%%%%%%%

\bibliographystyle{mnras}
\bibliography{martin} 

\begin{thebibliography}{}
\makeatletter
\relax
\def\mn@urlcharsother{\let\do\@makeother \do\$\do\&\do\#\do\^\do\_\do\%\do\~}
\def\mn@doi{\begingroup\mn@urlcharsother \@ifnextchar [ {\mn@doi@}
  {\mn@doi@[]}}
\def\mn@doi@[#1]#2{\def\@tempa{#1}\ifx\@tempa\@empty \href
  {http://dx.doi.org/#2} {doi:#2}\else \href {http://dx.doi.org/#2} {#1}\fi
  \endgroup}
\def\mn@eprint#1#2{\mn@eprint@#1:#2::\@nil}
\def\mn@eprint@arXiv#1{\href {http://arxiv.org/abs/#1} {{\tt arXiv:#1}}}
\def\mn@eprint@dblp#1{\href {http://dblp.uni-trier.de/rec/bibtex/#1.xml}
  {dblp:#1}}
\def\mn@eprint@#1:#2:#3:#4\@nil{\def\@tempa {#1}\def\@tempb {#2}\def\@tempc
  {#3}\ifx \@tempc \@empty \let \@tempc \@tempb \let \@tempb \@tempa \fi \ifx
  \@tempb \@empty \def\@tempb {arXiv}\fi \@ifundefined
  {mn@eprint@\@tempb}{\@tempb:\@tempc}{\expandafter \expandafter \csname
  mn@eprint@\@tempb\endcsname \expandafter{\@tempc}}}

\bibitem[\protect\citeauthoryear{{Allen}, {Evrard}  \& {Mantz}}{{Allen}
  et~al.}{2011}]{2011ARA&A..49..409A}
{Allen} S.~W.,  {Evrard} A.~E.,   {Mantz} A.~B.,  2011, \mn@doi [\araa]
  {10.1146/annurev-astro-081710-102514}, \href
  {https://ui.adsabs.harvard.edu/abs/2011ARA&A..49..409A} {49, 409}

\bibitem[\protect\citeauthoryear{{Applegate} et~al.,}{{Applegate}
  et~al.}{2014}]{2014MNRAS.439...48A}
{Applegate} D.~E.,  et~al., 2014, \mn@doi [\mnras] {10.1093/mnras/stt2129},
  \href {https://ui.adsabs.harvard.edu/abs/2014MNRAS.439...48A} {439, 48}

\bibitem[\protect\citeauthoryear{{Applegate} et~al.,}{{Applegate}
  et~al.}{2016}]{2016MNRAS.457.1522A}
{Applegate} D.~E.,  et~al., 2016, \mn@doi [\mnras] {10.1093/mnras/stw005},
  \href {https://ui.adsabs.harvard.edu/abs/2016MNRAS.457.1522A} {457, 1522}

\bibitem[\protect\citeauthoryear{{Bah{\'e}}, {McCarthy}  \& {King}}{{Bah{\'e}}
  et~al.}{2012}]{2012MNRAS.421.1073B}
{Bah{\'e}} Y.~M.,  {McCarthy} I.~G.,   {King} L.~J.,  2012, \mn@doi [\mnras]
  {10.1111/j.1365-2966.2011.20364.x}, \href
  {https://ui.adsabs.harvard.edu/abs/2012MNRAS.421.1073B} {421, 1073}

\bibitem[\protect\citeauthoryear{{Balm{\`e}s}, {Rasera}, {Corasaniti}  \&
  {Alimi}}{{Balm{\`e}s} et~al.}{2014}]{2014MNRAS.437.2328B}
{Balm{\`e}s} I.,  {Rasera} Y.,  {Corasaniti} P.~S.,   {Alimi} J.~M.,  2014,
  \mn@doi [\mnras] {10.1093/mnras/stt2050}, \href
  {https://ui.adsabs.harvard.edu/abs/2014MNRAS.437.2328B} {437, 2328}

\bibitem[\protect\citeauthoryear{{Bartelmann}}{{Bartelmann}}{1996}]{1996A&A...313..697B}
{Bartelmann} M.,  1996, \aap, \href
  {https://ui.adsabs.harvard.edu/abs/1996A%26A...313..697B} {313, 697}

\bibitem[\protect\citeauthoryear{{Becker} \& {Kravtsov}}{{Becker} \&
  {Kravtsov}}{2011}]{2011ApJ...740...25B}
{Becker} M.~R.,  {Kravtsov} A.~V.,  2011, \mn@doi [\apj]
  {10.1088/0004-637X/740/1/25}, \href
  {https://ui.adsabs.harvard.edu/abs/2011ApJ...740...25B} {740, 25}

\bibitem[\protect\citeauthoryear{{Bhattacharya}, {Habib}, {Heitmann}  \&
  {Vikhlinin}}{{Bhattacharya} et~al.}{2013}]{2013ApJ...766...32B}
{Bhattacharya} S.,  {Habib} S.,  {Heitmann} K.,   {Vikhlinin} A.,  2013,
  \mn@doi [\apj] {10.1088/0004-637X/766/1/32}, \href
  {https://ui.adsabs.harvard.edu/abs/2013ApJ...766...32B} {766, 32}

\bibitem[\protect\citeauthoryear{{Bocquet} et~al.,}{{Bocquet}
  et~al.}{2019}]{2019ApJ...878...55B}
{Bocquet} S.,  et~al., 2019, \mn@doi [\apj] {10.3847/1538-4357/ab1f10}, \href
  {https://ui.adsabs.harvard.edu/abs/2019ApJ...878...55B} {878, 55}

\bibitem[\protect\citeauthoryear{{Bullock}, {Kolatt}, {Sigad}, {Somerville},
  {Kravtsov}, {Klypin}, {Primack}  \& {Dekel}}{{Bullock}
  et~al.}{2001}]{2001MNRAS.321..559B}
{Bullock} J.~S.,  {Kolatt} T.~S.,  {Sigad} Y.,  {Somerville} R.~S.,  {Kravtsov}
  A.~V.,  {Klypin} A.~A.,  {Primack} J.~R.,   {Dekel} A.,  2001, \mn@doi
  [\mnras] {10.1046/j.1365-8711.2001.04068.x}, \href
  {https://ui.adsabs.harvard.edu/abs/2001MNRAS.321..559B} {321, 559}

\bibitem[\protect\citeauthoryear{{Child}, {Habib}, {Heitmann}, {Frontiere},
  {Finkel}, {Pope}  \& {Morozov}}{{Child} et~al.}{2018}]{2018ApJ...859...55C}
{Child} H.~L.,  {Habib} S.,  {Heitmann} K.,  {Frontiere} N.,  {Finkel} H.,
  {Pope} A.,   {Morozov} V.,  2018, \mn@doi [\apj] {10.3847/1538-4357/aabf95},
  \href {https://ui.adsabs.harvard.edu/abs/2018ApJ...859...55C} {859, 55}

\bibitem[\protect\citeauthoryear{{Chiu}, {Klein}, {Mohr}  \& {Bocquet}}{{Chiu}
  et~al.}{2023}]{2023MNRAS.tmp..949C}
{Chiu} I.~N.,  {Klein} M.,  {Mohr} J.,   {Bocquet} S.,  2023, \mn@doi [\mnras]
  {10.1093/mnras/stad957}, \href
  {https://ui.adsabs.harvard.edu/abs/2023MNRAS.tmp..949C} {}

\bibitem[\protect\citeauthoryear{{Diemer} \& {Joyce}}{{Diemer} \&
  {Joyce}}{2019}]{2019ApJ...871..168D}
{Diemer} B.,  {Joyce} M.,  2019, \mn@doi [\apj] {10.3847/1538-4357/aafad6},
  \href {https://ui.adsabs.harvard.edu/abs/2019ApJ...871..168D} {871, 168}

\bibitem[\protect\citeauthoryear{{Diemer} \& {Kravtsov}}{{Diemer} \&
  {Kravtsov}}{2014}]{2014ApJ...789....1D}
{Diemer} B.,  {Kravtsov} A.~V.,  2014, \mn@doi [\apj]
  {10.1088/0004-637X/789/1/1}, \href
  {https://ui.adsabs.harvard.edu/abs/2014ApJ...789....1D} {789, 1}

\bibitem[\protect\citeauthoryear{{Diemer} \& {Kravtsov}}{{Diemer} \&
  {Kravtsov}}{2015}]{2015ApJ...799..108D}
{Diemer} B.,  {Kravtsov} A.~V.,  2015, \mn@doi [\apj]
  {10.1088/0004-637X/799/1/108}, \href
  {https://ui.adsabs.harvard.edu/abs/2015ApJ...799..108D} {799, 108}

\bibitem[\protect\citeauthoryear{{Dietrich}, {B{\"o}hnert}, {Lombardi},
  {Hilbert}  \& {Hartlap}}{{Dietrich} et~al.}{2012}]{2012MNRAS.419.3547D}
{Dietrich} J.~P.,  {B{\"o}hnert} A.,  {Lombardi} M.,  {Hilbert} S.,   {Hartlap}
  J.,  2012, \mn@doi [\mnras] {10.1111/j.1365-2966.2011.19995.x}, \href
  {http://adsabs.harvard.edu/abs/2012MNRAS.419.3547D} {419, 3547}

\bibitem[\protect\citeauthoryear{{Dietrich} et~al.,}{{Dietrich}
  et~al.}{2019}]{2019MNRAS.483.2871D}
{Dietrich} J.~P.,  et~al., 2019, \mn@doi [\mnras] {10.1093/mnras/sty3088},
  \href {https://ui.adsabs.harvard.edu/abs/2019MNRAS.483.2871D} {483, 2871}

\bibitem[\protect\citeauthoryear{{Dolag}, {Komatsu}  \& {Sunyaev}}{{Dolag}
  et~al.}{2016}]{2016MNRAS.463.1797D}
{Dolag} K.,  {Komatsu} E.,   {Sunyaev} R.,  2016, \mn@doi [\mnras]
  {10.1093/mnras/stw2035}, \href
  {https://ui.adsabs.harvard.edu/abs/2016MNRAS.463.1797D} {463, 1797}

\bibitem[\protect\citeauthoryear{{Duffy}, {Schaye}, {Kay}  \& {Dalla
  Vecchia}}{{Duffy} et~al.}{2008}]{2008MNRAS.390L..64D}
{Duffy} A.~R.,  {Schaye} J.,  {Kay} S.~T.,   {Dalla Vecchia} C.,  2008, \mn@doi
  [\mnras] {10.1111/j.1745-3933.2008.00537.x}, \href
  {https://ui.adsabs.harvard.edu/abs/2008MNRAS.390L..64D} {390, L64}

\bibitem[\protect\citeauthoryear{{Dutton} \& {Macci{\`o}}}{{Dutton} \&
  {Macci{\`o}}}{2014}]{2014MNRAS.441.3359D}
{Dutton} A.~A.,  {Macci{\`o}} A.~V.,  2014, \mn@doi [\mnras]
  {10.1093/mnras/stu742}, \href
  {https://ui.adsabs.harvard.edu/abs/2014MNRAS.441.3359D} {441, 3359}

\bibitem[\protect\citeauthoryear{{Faltenbacher}, {Gottl{\"o}ber}, {Kerscher}
  \& {M{\"u}ller}}{{Faltenbacher} et~al.}{2002}]{2002A&A...395....1F}
{Faltenbacher} A.,  {Gottl{\"o}ber} S.,  {Kerscher} M.,   {M{\"u}ller} V.,
  2002, \mn@doi [\aap] {10.1051/0004-6361:20021263}, \href
  {https://ui.adsabs.harvard.edu/abs/2002A&A...395....1F} {395, 1}

\bibitem[\protect\citeauthoryear{{George} et~al.,}{{George}
  et~al.}{2012}]{2012ApJ...757....2G}
{George} M.~R.,  et~al., 2012, \mn@doi [\apj] {10.1088/0004-637X/757/1/2},
  \href {https://ui.adsabs.harvard.edu/abs/2012ApJ...757....2G} {757, 2}

\bibitem[\protect\citeauthoryear{{Gorenstein}, {Falco}  \&
  {Shapiro}}{{Gorenstein} et~al.}{1988}]{1988ApJ...327..693G}
{Gorenstein} M.~V.,  {Falco} E.~E.,   {Shapiro} I.~I.,  1988, \mn@doi [\apj]
  {10.1086/166226}, \href
  {https://ui.adsabs.harvard.edu/abs/1988ApJ...327..693G} {327, 693}

\bibitem[\protect\citeauthoryear{{Grandis}, {Mohr}, {Dietrich}, {Bocquet},
  {Saro}, {Klein}, {Paulus}  \& {Capasso}}{{Grandis}
  et~al.}{2019}]{2019MNRAS.488.2041G}
{Grandis} S.,  {Mohr} J.~J.,  {Dietrich} J.~P.,  {Bocquet} S.,  {Saro} A.,
  {Klein} M.,  {Paulus} M.,   {Capasso} R.,  2019, \mn@doi [\mnras]
  {10.1093/mnras/stz1778}, \href
  {https://ui.adsabs.harvard.edu/abs/2019MNRAS.488.2041G} {488, 2041}

\bibitem[\protect\citeauthoryear{{Grandis}, {Bocquet}, {Mohr}, {Klein}  \&
  {Dolag}}{{Grandis} et~al.}{2021}]{2021MNRAS.507.5671G}
{Grandis} S.,  {Bocquet} S.,  {Mohr} J.~J.,  {Klein} M.,   {Dolag} K.,  2021,
  \mn@doi [\mnras] {10.1093/mnras/stab2414}, \href
  {https://ui.adsabs.harvard.edu/abs/2021MNRAS.507.5671G} {507, 5671}

\bibitem[\protect\citeauthoryear{{Gupta}, {Saro}, {Mohr}, {Dolag}  \&
  {Liu}}{{Gupta} et~al.}{2017}]{2017MNRAS.469.3069G}
{Gupta} N.,  {Saro} A.,  {Mohr} J.~J.,  {Dolag} K.,   {Liu} J.,  2017, \mn@doi
  [\mnras] {10.1093/mnras/stx715}, \href
  {https://ui.adsabs.harvard.edu/abs/2017MNRAS.469.3069G} {469, 3069}

\bibitem[\protect\citeauthoryear{{Haiman}, {Mohr}  \& {Holder}}{{Haiman}
  et~al.}{2001}]{2001ApJ...553..545H}
{Haiman} Z.,  {Mohr} J.~J.,   {Holder} G.~P.,  2001, \mn@doi [\apj]
  {10.1086/320939}, \href
  {https://ui.adsabs.harvard.edu/abs/2001ApJ...553..545H} {553, 545}

\bibitem[\protect\citeauthoryear{{Henson}, {Barnes}, {Kay}, {McCarthy}  \&
  {Schaye}}{{Henson} et~al.}{2017}]{2017MNRAS.465.3361H}
{Henson} M.~A.,  {Barnes} D.~J.,  {Kay} S.~T.,  {McCarthy} I.~G.,   {Schaye}
  J.,  2017, \mn@doi [\mnras] {10.1093/mnras/stw2899}, \href
  {https://ui.adsabs.harvard.edu/abs/2017MNRAS.465.3361H} {465, 3361}

\bibitem[\protect\citeauthoryear{{Hopkins}, {Bahcall}  \& {Bode}}{{Hopkins}
  et~al.}{2005}]{2005ApJ...618....1H}
{Hopkins} P.~F.,  {Bahcall} N.~A.,   {Bode} P.,  2005, \mn@doi [\apj]
  {10.1086/425993}, \href
  {https://ui.adsabs.harvard.edu/abs/2005ApJ...618....1H} {618, 1}

\bibitem[\protect\citeauthoryear{{Ivezi{\'c}} et~al.,}{{Ivezi{\'c}}
  et~al.}{2019}]{2019ApJ...873..111I}
{Ivezi{\'c}} {\v{Z}}.,  et~al., 2019, \mn@doi [\apj]
  {10.3847/1538-4357/ab042c}, \href
  {https://ui.adsabs.harvard.edu/abs/2019ApJ...873..111I} {873, 111}

\bibitem[\protect\citeauthoryear{{Kaiser}, {Squires}  \& {Broadhurst}}{{Kaiser}
  et~al.}{1995}]{1995ApJ...449..460K}
{Kaiser} N.,  {Squires} G.,   {Broadhurst} T.,  1995, \mn@doi [\apj]
  {10.1086/176071}, \href
  {https://ui.adsabs.harvard.edu/abs/1995ApJ...449..460K} {449, 460}

\bibitem[\protect\citeauthoryear{{Kilbinger}}{{Kilbinger}}{2015}]{2015RPPh...78h6901K}
{Kilbinger} M.,  2015, \mn@doi [Reports on Progress in Physics]
  {10.1088/0034-4885/78/8/086901}, \href
  {https://ui.adsabs.harvard.edu/abs/2015RPPh...78h6901K} {78, 086901}

\bibitem[\protect\citeauthoryear{{Klypin}, {Yepes}, {Gottl{\"o}ber}, {Prada}
  \& {He{\ss}}}{{Klypin} et~al.}{2016}]{2016MNRAS.457.4340K}
{Klypin} A.,  {Yepes} G.,  {Gottl{\"o}ber} S.,  {Prada} F.,   {He{\ss}} S.,
  2016, \mn@doi [\mnras] {10.1093/mnras/stw248}, \href
  {https://ui.adsabs.harvard.edu/abs/2016MNRAS.457.4340K} {457, 4340}

\bibitem[\protect\citeauthoryear{{Laureijs} et~al.,}{{Laureijs}
  et~al.}{2011}]{2011arXiv1110.3193L}
{Laureijs} R.,  et~al., 2011, arXiv:1110.3193

\bibitem[\protect\citeauthoryear{{Lee}, {Le Brun}, {Haq}, {Deering}, {King},
  {Applegate}  \& {McCarthy}}{{Lee} et~al.}{2018}]{2018MNRAS.479..890L}
{Lee} B.~E.,  {Le Brun} A.~M.~C.,  {Haq} M.~E.,  {Deering} N.~J.,  {King}
  L.~J.,  {Applegate} D.,   {McCarthy} I.~G.,  2018, \mn@doi [\mnras]
  {10.1093/mnras/sty1377}, \href
  {https://ui.adsabs.harvard.edu/abs/2018MNRAS.479..890L} {479, 890}

\bibitem[\protect\citeauthoryear{{Ludlow}, {Navarro}, {Angulo},
  {Boylan-Kolchin}, {Springel}, {Frenk}  \& {White}}{{Ludlow}
  et~al.}{2014}]{2014MNRAS.441..378L}
{Ludlow} A.~D.,  {Navarro} J.~F.,  {Angulo} R.~E.,  {Boylan-Kolchin} M.,
  {Springel} V.,  {Frenk} C.,   {White} S. D.~M.,  2014, \mn@doi [\mnras]
  {10.1093/mnras/stu483}, \href
  {https://ui.adsabs.harvard.edu/abs/2014MNRAS.441..378L} {441, 378}

\bibitem[\protect\citeauthoryear{{Ludlow}, {Bose}, {Angulo}, {Wang},
  {Hellwing}, {Navarro}, {Cole}  \& {Frenk}}{{Ludlow}
  et~al.}{2016}]{2016MNRAS.460.1214L}
{Ludlow} A.~D.,  {Bose} S.,  {Angulo} R.~E.,  {Wang} L.,  {Hellwing} W.~A.,
  {Navarro} J.~F.,  {Cole} S.,   {Frenk} C.~S.,  2016, \mn@doi [\mnras]
  {10.1093/mnras/stw1046}, \href
  {https://ui.adsabs.harvard.edu/abs/2016MNRAS.460.1214L} {460, 1214}

\bibitem[\protect\citeauthoryear{{Mandelbaum}, {Tasitsiomi}, {Seljak},
  {Kravtsov}  \& {Wechsler}}{{Mandelbaum} et~al.}{2005}]{2005MNRAS.362.1451M}
{Mandelbaum} R.,  {Tasitsiomi} A.,  {Seljak} U.,  {Kravtsov} A.~V.,
  {Wechsler} R.~H.,  2005, \mn@doi [\mnras] {10.1111/j.1365-2966.2005.09417.x},
  \href {https://ui.adsabs.harvard.edu/abs/2005MNRAS.362.1451M} {362, 1451}

\bibitem[\protect\citeauthoryear{{Mantz}, {Allen}, {Morris}, {Rapetti},
  {Applegate}, {Kelly}, {von der Linden}  \& {Schmidt}}{{Mantz}
  et~al.}{2014}]{2014MNRAS.440.2077M}
{Mantz} A.~B.,  {Allen} S.~W.,  {Morris} R.~G.,  {Rapetti} D.~A.,  {Applegate}
  D.~E.,  {Kelly} P.~L.,  {von der Linden} A.,   {Schmidt} R.~W.,  2014,
  \mn@doi [\mnras] {10.1093/mnras/stu368}, \href
  {https://ui.adsabs.harvard.edu/abs/2014MNRAS.440.2077M} {440, 2077}

\bibitem[\protect\citeauthoryear{{McClintock} et~al.,}{{McClintock}
  et~al.}{2019}]{2019MNRAS.482.1352M}
{McClintock} T.,  et~al., 2019, \mn@doi [\mnras] {10.1093/mnras/sty2711}, \href
  {https://ui.adsabs.harvard.edu/abs/2019MNRAS.482.1352M} {482, 1352}

\bibitem[\protect\citeauthoryear{{Meneghetti} et~al.,}{{Meneghetti}
  et~al.}{2014}]{2014ApJ...797...34M}
{Meneghetti} M.,  et~al., 2014, \mn@doi [\apj] {10.1088/0004-637X/797/1/34},
  \href {https://ui.adsabs.harvard.edu/abs/2014ApJ...797...34M} {797, 34}

\bibitem[\protect\citeauthoryear{{Navarro}, {Frenk}  \& {White}}{{Navarro}
  et~al.}{1997}]{1997ApJ...490..493N}
{Navarro} J.~F.,  {Frenk} C.~S.,   {White} S. D.~M.,  1997, \mn@doi [\apj]
  {10.1086/304888}, \href
  {https://ui.adsabs.harvard.edu/abs/1997ApJ...490..493N} {490, 493}

\bibitem[\protect\citeauthoryear{{Oguri} \& {Hamana}}{{Oguri} \&
  {Hamana}}{2011}]{2011MNRAS.414.1851O}
{Oguri} M.,  {Hamana} T.,  2011, \mn@doi [\mnras]
  {10.1111/j.1365-2966.2011.18481.x}, \href
  {https://ui.adsabs.harvard.edu/abs/2011MNRAS.414.1851O} {414, 1851}

\bibitem[\protect\citeauthoryear{{Planck Collaboration} et~al.,}{{Planck
  Collaboration} et~al.}{2016}]{2016A&A...594A..24P}
{Planck Collaboration} et~al., 2016, \mn@doi [\aap]
  {10.1051/0004-6361/201525833}, \href
  {https://ui.adsabs.harvard.edu/abs/2016A&A...594A..24P} {594, A24}

\bibitem[\protect\citeauthoryear{{Prada}, {Klypin}, {Cuesta}, {Betancort-Rijo}
  \& {Primack}}{{Prada} et~al.}{2012}]{2012MNRAS.423.3018P}
{Prada} F.,  {Klypin} A.~A.,  {Cuesta} A.~J.,  {Betancort-Rijo} J.~E.,
  {Primack} J.,  2012, \mn@doi [\mnras] {10.1111/j.1365-2966.2012.21007.x},
  \href {https://ui.adsabs.harvard.edu/abs/2012MNRAS.423.3018P} {423, 3018}

\bibitem[\protect\citeauthoryear{{Ragagnin}, {Dolag}, {Moscardini}, {Biviano}
  \& {D'Onofrio}}{{Ragagnin} et~al.}{2019}]{2019MNRAS.486.4001R}
{Ragagnin} A.,  {Dolag} K.,  {Moscardini} L.,  {Biviano} A.,   {D'Onofrio} M.,
  2019, \mn@doi [\mnras] {10.1093/mnras/stz1103}, \href
  {https://ui.adsabs.harvard.edu/abs/2019MNRAS.486.4001R} {486, 4001}

\bibitem[\protect\citeauthoryear{{Ramdas}, {Garcia}  \& {Cuturi}}{{Ramdas}
  et~al.}{2015}]{2015arXiv150902237R}
{Ramdas} A.,  {Garcia} N.,   {Cuturi} M.,  2015, arXiv e-prints, \href
  {https://ui.adsabs.harvard.edu/abs/2015arXiv150902237R} {p. arXiv:1509.02237}

\bibitem[\protect\citeauthoryear{{Schneider} \& {Seitz}}{{Schneider} \&
  {Seitz}}{1995}]{1995A&A...294..411S}
{Schneider} P.,  {Seitz} C.,  1995, \aap, \href
  {https://ui.adsabs.harvard.edu/abs/1995A&A...294..411S} {294, 411}

\bibitem[\protect\citeauthoryear{{Schrabback} et~al.,}{{Schrabback}
  et~al.}{2018}]{2018MNRAS.474.2635S}
{Schrabback} T.,  et~al., 2018, \mn@doi [\mnras] {10.1093/mnras/stx2666}, \href
  {http://adsabs.harvard.edu/abs/2018MNRAS.474.2635S} {474, 2635}

\bibitem[\protect\citeauthoryear{{Schrabback} et~al.,}{{Schrabback}
  et~al.}{2021}]{2021MNRAS.505.3923S}
{Schrabback} T.,  et~al., 2021, \mn@doi [\mnras] {10.1093/mnras/stab1386},
  \href {https://ui.adsabs.harvard.edu/abs/2021MNRAS.505.3923S} {505, 3923}

\bibitem[\protect\citeauthoryear{{Seitz} \& {Schneider}}{{Seitz} \&
  {Schneider}}{1997}]{1997A&A...318..687S}
{Seitz} C.,  {Schneider} P.,  1997, \aap, \href
  {https://ui.adsabs.harvard.edu/abs/1997A&A...318..687S} {318, 687}

\bibitem[\protect\citeauthoryear{{Seljak}}{{Seljak}}{2000}]{2000MNRAS.318..203S}
{Seljak} U.,  2000, \mn@doi [\mnras] {10.1046/j.1365-8711.2000.03715.x}, \href
  {https://ui.adsabs.harvard.edu/abs/2000MNRAS.318..203S} {318, 203}

\bibitem[\protect\citeauthoryear{{Shan} et~al.,}{{Shan}
  et~al.}{2017}]{2017ApJ...840..104S}
{Shan} H.,  et~al., 2017, \mn@doi [\apj] {10.3847/1538-4357/aa6c68}, \href
  {https://ui.adsabs.harvard.edu/abs/2017ApJ...840..104S} {840, 104}

\bibitem[\protect\citeauthoryear{{Sommer}, {Schrabback}, {Applegate},
  {Hilbert}, {Ansarinejad}, {Floyd}  \& {Grandis}}{{Sommer}
  et~al.}{2022}]{2022MNRAS.509.1127S}
{Sommer} M.~W.,  {Schrabback} T.,  {Applegate} D.~E.,  {Hilbert} S.,
  {Ansarinejad} B.,  {Floyd} B.,   {Grandis} S.,  2022, \mn@doi [\mnras]
  {10.1093/mnras/stab3052}, \href
  {https://ui.adsabs.harvard.edu/abs/2022MNRAS.509.1127S} {509, 1127}

\bibitem[\protect\citeauthoryear{{Staniszewski} et~al.,}{{Staniszewski}
  et~al.}{2009}]{2009ApJ...701...32S}
{Staniszewski} Z.,  et~al., 2009, \mn@doi [\apj] {10.1088/0004-637X/701/1/32},
  \href {http://adsabs.harvard.edu/abs/2009ApJ...701...32S} {701, 32}

\bibitem[\protect\citeauthoryear{{Sunyaev} \& {Zeldovich}}{{Sunyaev} \&
  {Zeldovich}}{1970}]{1970CoASP...2...66S}
{Sunyaev} R.~A.,  {Zeldovich} Y.~B.,  1970, Comments on Astrophysics and Space
  Physics, \href {http://adsabs.harvard.edu/abs/1970CoASP...2...66S} {2, 66}

\bibitem[\protect\citeauthoryear{{Sunyaev} \& {Zeldovich}}{{Sunyaev} \&
  {Zeldovich}}{1980}]{1980ARA&A..18..537S}
{Sunyaev} R.~A.,  {Zeldovich} I.~B.,  1980, \mn@doi [\araa]
  {10.1146/annurev.aa.18.090180.002541}, \href
  {http://adsabs.harvard.edu/abs/1980ARA%26A..18..537S} {18, 537}

\bibitem[\protect\citeauthoryear{Tavio, Cuesta, Prada, Klypin  \&
  Sanchez-Conde}{Tavio et~al.}{2008}]{2008arXiv0807.3027T}
Tavio H.,  Cuesta A.~J.,  Prada F.,  Klypin A.~A.,   Sanchez-Conde M.~A.,
  2008, arXiv:2103.16212

\bibitem[\protect\citeauthoryear{{Tollet} et~al.,}{{Tollet}
  et~al.}{2016}]{2016MNRAS.456.3542T}
{Tollet} E.,  et~al., 2016, \mn@doi [\mnras] {10.1093/mnras/stv2856}, \href
  {https://ui.adsabs.harvard.edu/abs/2016MNRAS.456.3542T} {456, 3542}

\bibitem[\protect\citeauthoryear{Villani}{Villani}{2008}]{villani2008}
Villani C.,  2008, Optimal transport -- Old and new.
Springer Verlag, pp xxii+973, \mn@doi{10.1007/978-3-540-71050-9}

\bibitem[\protect\citeauthoryear{{Wright} \& {Brainerd}}{{Wright} \&
  {Brainerd}}{2000}]{2000ApJ...534...34W}
{Wright} C.~O.,  {Brainerd} T.~G.,  2000, \mn@doi [\apj] {10.1086/308744},
  \href {https://ui.adsabs.harvard.edu/abs/2000ApJ...534...34W} {534, 34}

\bibitem[\protect\citeauthoryear{{Zhang}, {Yang}, {Faltenbacher}, {Springel},
  {Lin}  \& {Wang}}{{Zhang} et~al.}{2009}]{2009ApJ...706..747Z}
{Zhang} Y.,  {Yang} X.,  {Faltenbacher} A.,  {Springel} V.,  {Lin} W.,   {Wang}
  H.,  2009, \mn@doi [\apj] {10.1088/0004-637X/706/1/747}, \href
  {https://ui.adsabs.harvard.edu/abs/2009ApJ...706..747Z} {706, 747}

\bibitem[\protect\citeauthoryear{{Zohren} et~al.,}{{Zohren}
  et~al.}{2022}]{2022A&A...668A..18Z}
{Zohren} H.,  et~al., 2022, \mn@doi [\aap] {10.1051/0004-6361/202142991}, \href
  {https://ui.adsabs.harvard.edu/abs/2022A&A...668A..18Z} {668, A18}

\makeatother
\end{thebibliography}

%%%%%%%%%%%%%%%%% APPENDICES %%%%%%%%%%%%%%%%%%%%%

\appendix

\section{Notation used in this paper}
\label{sec:appendix:notation}

In \tabl~\ref{tab:notation} we give an overview of symbols and notation specific to this work. 

\begin{table*}
\centering
\begin{tabular}{r c l} 
 \hline
 Designation & Defined in \sect & Description \\
 \hline
 $\rmis$ & \ref{sec:meth:miscdistr} & Distance in the sky plane between a measured center and the bottom of the \\
 & & ~gravitational potential (miscentring radius) \\
 $\meanrmis$ & \ref{sec:meth:miscdistr} & Mean of $\rmis$, given a miscentring distribution \\
 $\rmin$ & \ref{sec:meth:radialmodel} & Minimum radius for fitting masses from reduced shear \\
 $\rmax$ & \ref{sec:meth:radialmodel} & Maximum radius for fitting masses from reduced shear \\
 $R$     & \ref{sec:meth:miscmodel} & Distance from gravitational center to a point at which we measure reduced shear \\
 $\xi$   & \ref{sec:meth:miscmodel} & Distance from miscentered position to a point at which we measure reduced shear  \\
 $\cfit$ & \ref{sec:meth:miscmodel} & Denotes centered mass fitting from reduced shear, or that a lensing quantity is\\ 
  &  & ~taken with respect to the true center\\
 $\mfit$ & \ref{sec:meth:miscmodel} & Denotes miscentered mass fitting from reduced shear, or that a lensing quantity\\
  &  & ~is taken with respect to the observed center\\
 $\truemass$   & \ref{sec:meth:massbias} & True mass of an individual halo \\
 $\wlmass$   & \ref{sec:meth:massbias} & Weak lensing mass of an individual halo \\
 $\bias$  & \ref{sec:meth:massbias} & Weak lensing mass bias \\
 $\logbias$ & \ref{sec:meth:massbias} & Natural logarithm of the weak lensing bias \\
 $\munormal$ & \ref{sec:meth:massbias} & Estimated mean of $\bias$ \\
 $\sigmanormal$ & \ref{sec:meth:massbias} & Estimated standard deviation of $\bias$ \\
 $\mulognormal$ & \ref{sec:meth:massbias} & Estimated mean of $\logbias$ \\
 $\sigmalognormal$ & \ref{sec:meth:massbias} & Estimated standard deviation of $\logbias$ \\
 $\nonrandommiscentring$ & \ref{sec:meth:overcorr} & Denotes non-random miscentring \\
 $\randommiscentring$ & \ref{sec:meth:overcorr} & Denotes randomized miscentring \\ 
 $\overcorrectionnormal$ & \ref{sec:meth:overcorr} & Mean mass over-correction factor assuming a Gaussian bias distribution \\
 $\overcorrectionlognormal$ &  \ref{sec:meth:overcorr} & Mean mass over-correction factor assuming a log-normal bias distribution \\
 $\wnormal$ & \ref{sec:meth:wasserstein} & Estimated deviation (in standard deviations) from a normal distribution of $\bias$ \\
 $\wlognormal$ & \ref{sec:meth:wasserstein} & Estimated deviation (in standard deviations) from a normal distribution of $\logbias$ \\
\end{tabular}
\caption{Overview of non-standard notations used in this work. }
\label{tab:notation}
\end{table*}

%%%%%%%%%%%%%%%%%%%%%%%%%%%%%%%%%%%%%%%%%%%%%%%%%%

% Don't change these lines
\bsp	% typesetting comment
\label{lastpage}
\end{document}